\documentclass[aps,prd,reprint,showpacs,floatfix,longbibliography,,superscriptaddress]{revtex4-1}
\usepackage{tikz,graphicx,booktabs,multirow,array,microtype,mathtools,float}
\usepackage{enumerate}   
\extrafloats{100}
\usepackage[colorlinks=true,backref=false, linktocpage=true,
citecolor=blue,
urlcolor=blue,linkcolor=blue,
pdfpagemode=UseOutlines]{hyperref}
\newcolumntype{C}{>{$}c<{$}}
\usepackage{dcolumn}
\AtBeginDocument{
\heavyrulewidth=.08em
\lightrulewidth=.05em
\cmidrulewidth=.03em
\belowrulesep=.65ex
\belowbottomsep=0pt
\aboverulesep=.4ex
\abovetopsep=0pt
\cmidrulesep=\doublerulesep
\cmidrulekern=.5em
\defaultaddspace=.5em
}

\usepackage{amsmath} 
\usepackage{dsfont}  
\usepackage{bm}      
\def\mw{\mathcal{W}}
\def\mm{\mathcal{M}}
\def\iden{\mathds{1}}
\def\reals{\mathds{R}}
\def\beq{\begin{equation}}
\def\eeq{\end{equation}}
\def\beqs#1\eeqs{\beq\begin{split} #1 \end{split}\eeq}

\long\def\comment#1{}

\begin{document}
\title{Scattering phaseshift formulas for mesons and baryons in elongated boxes}
\author{Frank X. Lee}
\affiliation{Physics Department, The George Washington University, Washington, DC 20052, USA}
\email{fxlee@gwu.edu}
\author{Andrei Alexandru}
\email{aalexan@gwu.edu}
\affiliation{Physics Department, The George Washington University, Washington, DC 20052, USA}
\affiliation{Albert Einstein Center for Fundamental Physics, Institute for Theoretical Physics, University of Bern, Sidlerstrasse 5, CH-3012 Bern, Switzerland}
\date{\today}

\begin{abstract}
We derive L\"{u}scher phaseshift formulas for two-particle states in boxes elongated in one of the dimensions. 
Such boxes offer a cost-effective way of varying the relative momentum of the particles. 
Boosted states in the elongated direction, which allow wider access to energies, are also considered. 
The formulas for the various scenarios 
(moving and zero-momentum states in cubic and elongated boxes) are compared 
and relations between them are clarified.
The results are applicable to a wide set of meson-meson and meson-baryon elastic scattering processes, with the two-particle system having equal or unequal masses.
\end{abstract} 


\maketitle

\section{Introduction}
Hadron-hadron scattering is an indispensable  tool in understanding the nature of the strong nuclear force, 
both experimentally and theoretically.
The theoretical groundwork was laid out by L\"{u}scher~\cite{Luscher:1990ux} who 
showed how to relate elastic scattering phaseshifts with the energies of the two-body states in a finite box. 
Various extensions to the method have since been made to enhance its applications, 
including moving frames~\cite{Rummukainen:1995vs}, 
moving frame involving unequal masses and baryons~\cite{Fu:2011xz,Leskovec:2012gb,Gockeler:2012yj},
asymmetric boxes~\cite{Feng:2004ua}, and more recently inelastic scattering~\cite{Li:2012bi,Briceno:2014oea}.
The use of asymmetric boxes has proven to be efficient in recent studies of the $\rho$ meson 
resonance in $\pi\pi$ scattering~\cite{Guo:2016zos,Pelissier:2012pi}.
Instead of varying the size of the entire box, only one side is elongated, requiring much less computing resources.
Our main goal in this work is to derive the phaseshift formulas needed to study meson-baryon elastic  scattering in elongated boxes, 
with an eye towards a lattice QCD simulation of  the $\Delta$ resonance in $\pi N$ scattering.

\section{Phaseshift formalism}

In infinite volume, standard quantum mechanics defines elastic scattering phaseshift as the change in phase 
in the scattered wave relative to the incident wave in the asymptotic region where the interaction can be neglected.
In the partial-wave expansion, the wavefunction is 
$\psi(\bm r) = e^{ikz} + f(\theta) {e^{ikr} \over r}$ where
$f(\theta)=\sum_{l=0}^\infty (2l+1)f_lP_l(\cos\theta)$ is the scattering amplitude, and phaseshift $\delta_l$ enters via $f_l={e^{2i\delta_l}-1 \over 2 i k}$.
The phaseshift is a real-valued function of the interaction energy and carries information about the nature of the interaction, 
such as whether the force is 
attractive ($\delta<0$) or repulsive ($\delta>0$), whether a resonance is formed in the scattering, etc. 
 Scattering length can also be extracted in its effective range expansion.  
The phaseshift can be determined in the region where the interaction is vanishing, so the solution to the 
Schr\"{o}dinger equation (which has the form of a Helmholtz equation)
\beq
(\nabla^2 + k^2) \psi(\bm r)=0\,,
\label{eq:hemholtz}
\eeq
 can be expressed in terms of ordinary spherical bessel functions
$ \psi \propto [   \alpha_l j_l(kr) + \beta_l n_l(kr) ]$ where the coefficients can be found by 
matching up with the wavefunction in the interior.
The phaseshift can be computed from the coefficients by
\beq
e^{2i\delta_l(k)} = {\alpha_l(k) + i \beta_l(k) \over \alpha_l(k) - i \beta_l(k)  }\,.
\eeq

On the lattice, a similar procedure can be followed to study scattering phaseshifts as shown by L\"{u}scher, except that the system is 
now confined in a box of size $L$ (we assume the size is big enough so that  the interaction range $R<L/2$). 
The wavefunction must satisfy periodic boundary conditions 
\beq \psi(\bm r + \bm n L) =\psi(\bm r )\eeq
 so the solutions to 
the Helmholtz equation are zeta functions instead of Bessel functions. 
Basically, one ends up with a new relation that connects the phaseshifts with the
energies of two-body states in the box, 
in the form of a matrix equation~\cite{Luscher:1990ux}
\beq
\det \left [e^{2i\delta(k,L)} - {\mathcal{M}(k,L) + i \over \mathcal{M}(k,L) -i} \right ] =0\,,
\label{eq:phaselat}
\eeq
for positive relative momentum $k$. The $\mathcal{M}(k,L)$ is a well-defined matrix in terms of zeta functions,  
and it is purely a mathematical function of the relative momentum and box geometry.

The L\"{u}scher method provides a general strategy to understand hadron resonances via the phaseshift from the first principles of QCD. 
The basic idea is to take advantage of the dependence of the phaseshift $\delta(k,L)$ on momentum $k$ and box size $L$. 
The interaction energies can be computed on different boxes, thus  
allowing access to the phaseshift via the ``L\"{u}scher formula'' afforded by Eq.~\ref{eq:phaselat}.
Two comments are in order about the method. 
1) In a box, the matching of wavefunctions on the interaction boundary (to 
obtain the coefficients $\alpha_l$ and $\beta_l$) is replaced by the matrix function $\mathcal{M}(k,L)$. 
The nature of the interaction is encoded in the energies of the two-body states, which is measured through the quark-gluon dynamics of QCD.
2)  Even though the energies are obtained in Euclidean time in a finite box, 
the phaseshift has the same definition and meaning as in infinite volume and Minkowski time. 
Once the pion mass is brought to the physical point,  the phaseshift computed in the box can be directly compared with experiment.
The box is simply a device that serves two purposes at once: 
to allow the interaction energy to be computed in lattice QCD by making the problem finite thus
amenable to a numerical approach, and to facilitate the access to the physical phaseshift via the L\"{u}scher method.

\begin{widetext}

\section{Phaseshift reduction in the elongated box }

The phaseshift formula in Eq.~\ref{eq:phaselat} must be adapted to the symmetry imposed by the box. 
The issue arises because symmetries in the infinite volume are reduced to the symmetries in the box. 
Internal symmetries like color, flavor, and isospin are not affected. 
But angular momentum, which is the measure of rotational symmetries, is greatly affected.  
Specific to the scattering problem in the box,  two areas will be impacted; one is the phaseshift formula used to extract the phaseshifts, 
the other the interpolators used to construct the scattering states. Our focus is on the former; the latter will be addressed separately.
The interpolators must transform according to the symmetries in the box, not in the infinite volume. 
How the symmetry is reduced from the infinite volume to the periodic box and vice versa is a technical but important issue 
since we compute energies for two-body states in the box. The subject can be treated formally by group theory.

We consider a box elongated in the $z$-direction as illustrated in Fig.~\ref{fig:box_elongated}.
The infinite volume symmetry group for spatial rotations is $SO(3)$ which has an infinite number of elements 
and irreducible representations (irreps) labeled by angular momentum $J$. 
For the elongated box, the symmetry group is called the dihedral (or tetragonal) group $D_{4}$ which has 8 elements and 5 irreps.
$D_{4}$ is a finite subgroup of the rotation group $SO(3)$.
To include half-integral angular momentum as required for baryons,  
its double cover group $^2D_{4}$ is needed which has 16 elements and 7 irreps. 
(In infinite volume, the double-cover group of $SO(3)$ is $SU(2)$.)
For situations where parity is a good quantum number, the full symmetry group of the elongated box must include space inversion (parity). This group, denoted by $^2D_{4h}$, has 32 elements
and 14 irreps.  The full technical details of the $^2D_{4h}$ group are given in Appendix~\ref{app:elongated}.
\begin{figure}
\includegraphics[scale=0.5]{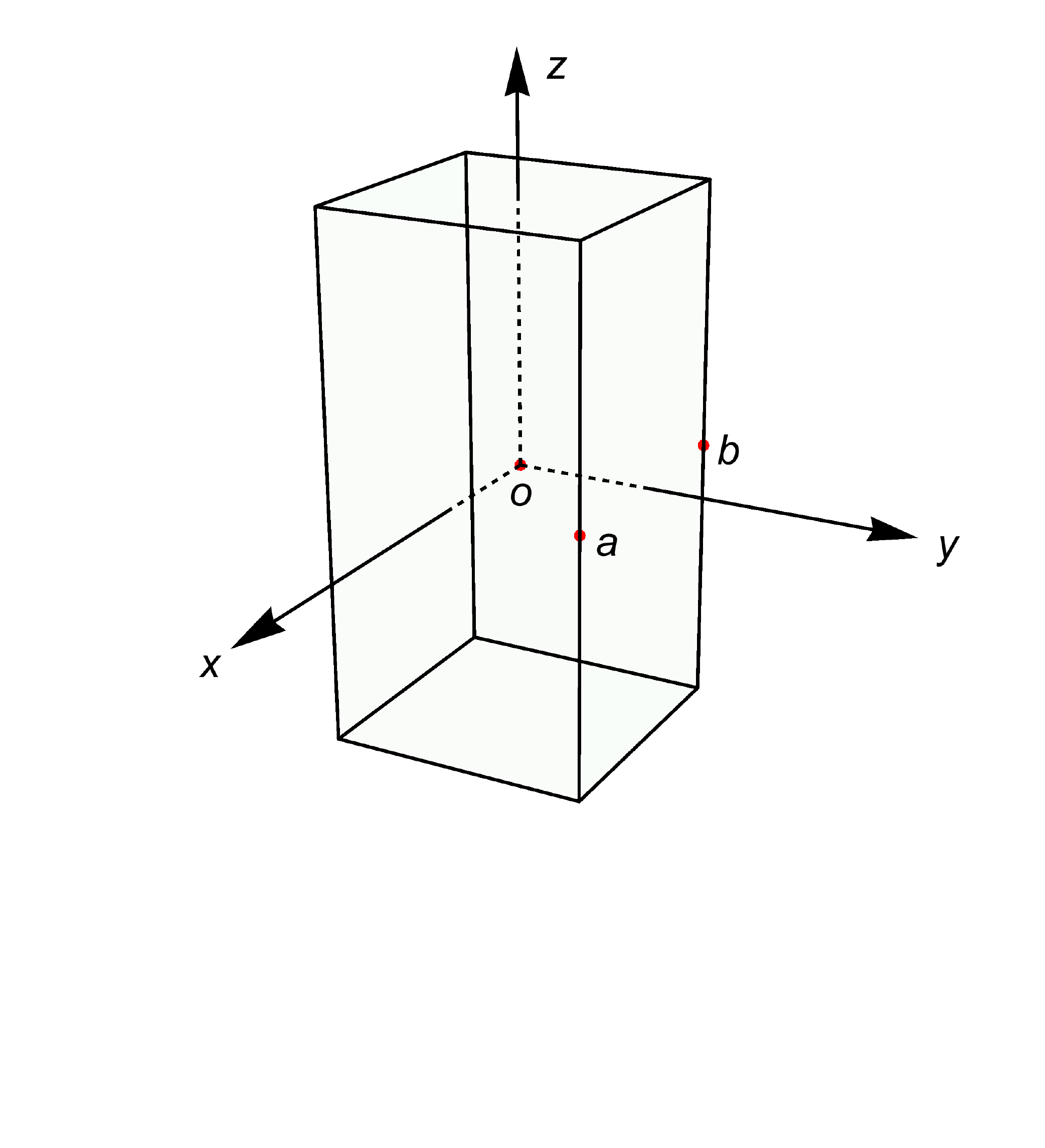}
\vspace{-3.5cm}
\caption{Symmetry operations that form the dihedral $D_4$ group in the elongated box whose dimensions are $L\times L \times \eta L$ where $\eta$ is the elongation factor in the $z$-direction. 
The group has 8 elements (rotations that leave the elongated box invariant), which can be divided into 5 conjugacy classes: 
the identity ($I$), two $\pi/2$ rotations about the $z$-axis, 
one $\pi$ rotation about the $z$-axis,  two $\pi$ rotations about $x$ and $y$ axes, and  two $\pi$  rotations about the two diagonals in the $xy$-plane denoted by $Oa$ and $Ob$. 
The operations are performed in a right-hand way with the thumb pointing from the center to the various symmetry points. 
The full details of the group, with the inclusion of half-integer spin and spatial inversion, 
are given in Appendix~\ref{app:elongated}.
}
\label{fig:box_elongated}
\end{figure}

In this paper we will be concerned with two-particle states in an elongated box. Our
goal is to derive the relevant L\"uscher formulas for the irreps of $^2D_{4h}$ corresponding
to meson-baryon states, where the meson is spinless and the baryon has spin
$1/2$. 
As a cross-check, we also derive  the relevant L\"uscher formulas for scattering
of spinless mesons.

In general, the total angular momentum for scattering two particles with spin $S_1$ and 
$S_2$ is 
\beq 
\bm J= \bm S_1 + \bm S_2 + \bm l \,, \label{eq:Jsum} 
\eeq 
where $l$ is the relative orbital angular momentum (partial-waves).
For the asymptotic states, when the particles are far away from each other, they are not 
interacting and we can label the states according to $S_i$ and $(S_i)_z$,
or equivalently with $S$, $S_i$, and $S_z$, where $\bm S=\bm S_1+\bm S_2$ is the total spin.
The scattering conserves $\bm J$, but can change both $\bm l$ and $\bm S$. 
For the cases considered in this
paper, one of the particles will be spinless so that total spin is simply the spin of the other
particle and the situation simplifies since there are no possible changes in the total spin. 
Moreover, the orbital angular momentum also remains fixed for our cases:
In the case that $S=0$ we have $l=J$ which is conserved. When $S=1/2$ for a given $J$, 
$l$ can assume two different values. These two channels have different parity and since
the parity is conserved the value of $l$ cannot change. Thus for $S=1/2$ we can use the 
parity of the state to identify which $l$ corresponds
to a given $J$. As such, in some of the formulas and tables in this paper we will
indicate the relevant channel by labeling $J$ and the parity.
Some comments on parity are in order here.
The total parity of the two particle state is equal to 
\beq P_{tot}=P_1P_2 (-1)^l \eeq 
where $P_1$ and $P_2$ are the intrinsic parities of the two scattering particles. For simplicity
we will assume that the intrinsic parity $P_1 P_2$ is positive. For the other case, the parity
assignments indicated in the tables and formulas will simply be reversed.

\subsection{Angular momentum in the elongated box}
\label{sec:ang_elongated}

For spherically symmetric interactions the eigenstates of the Hamiltonian in the
infinite volume form multiplets that furnish bases for the irreps of $SU(2)$, the double cover
of the rotations group. These multiplets are labeled by the angular momentum 
$J=0, {1\over 2}, 1, {3\over 2}, 2,\ldots$.
For elongated boxes, these multiplets split into smaller sets that mix under the action of rotations that leave the box invariant, forming the bases for one of the 7 irreps of the $^2D_{4}$ group. 
Then the question is: for a given $J$, what irreps are coupled to it? 
To answer this we have to decompose the irrep $J$ of the full rotation group $SU(2)$,   
into a direct sum of the irreps of the $^2D_{4}$ group,  $J=\bigoplus_i n(\Gamma_i, J) \Gamma_i$,  
where the coefficient is called the multiplicity, which tells how many times irrep $\Gamma_i$ appears in the given $J$.
This can be calculated (see for example~\cite{Johnson:1982, Moore:2005dw}) using 
\beq
n(\Gamma_i,J) = {1\over g} \sum_k n_k \chi(k,\Gamma_i) \,  \chi(\omega_k,J).
\label{eq:decom}
\eeq
The index $k$ runs through all  7 classes of $^2D_{4}$ and $g=16$ is the total number of elements in the group. 
$n_k$ is the number of elements in the $k$-th class, and $\chi(k,\Gamma_i)$ are the characters given in Table~\ref{tab:char2D4} in the appendix.
$\chi(\omega_k,J)$ stands for the character of full rotation group for angular momentum $J$ and rotation angle $\omega_k$ in class $k$.
This can be computed as follows~\cite{Tinkham:1992}.
Any rotation $k$ is characterized by a rotation axis and the rotation angle $\omega_k$. 
Since the character (trace) of the matrix is invariant under similarity transformations the result
will be equal to an equivalent rotation around the $z$-axis (the similarity matrix in this
case is simply a rotation that takes the rotation axis into the $z$-axis).
The character is then the trace of this diagonal matrix
\beq
\chi(\omega_k,J) =\sum_{m=-J}^{J} e^{-i m \omega_k} = {\sin[(J+1/2) \omega_k ]\over \sin(\omega_k/2)}.
\label{eq:charSU2}
\eeq
The results of the decomposition from applying Eq.~\ref{eq:decom} are given in Table~\ref{tab:2D4h}.
Note that limits must be taken if division by zero is encountered in evaluating Eq.~\ref{eq:charSU2}.

The parity transformation, $i$, is simply added to $^2D_{4}$ by taking the direct product,
$^2D_{4h}={^2D_{4}}\otimes \{I, i\}$, so that the irreps are simply doubled each one
generating a positive and negative parity irrep. To work out the decomposition of $J$
multiplet into irreps of $^2D_{4h}$ we assume that the states have parity $(-1)^l$
($l=J$ for integer $J$ and $l=J\pm1/2$ for half-integer cases).

The left part of  Table~\ref{tab:2D4h} shows that angular momenta $J= 0$ and $J=1/2$ 
correspond to single irreps $A^+_1$ and $G^\pm_1$, respectively, 
but it does not mean they have an one-to-one correspondence because 
the same irrep appears at higher $J$ values, sometimes multiple times in the same $J$. Then the question is:
If an energy eigenstate in the box belongs to one of the irreps, what angular momentum content does it have? 
In other words, what is the inverse of the correspondence displayed in the left half of the table?
If we restrict to $J\leq 9$, the result is shown in the right half of the table.
Assuming that the states with higher $J$-values have higher energy,
we see that it is reasonable to identify the ground state in the $A^+_1$ irrep as $J=0$, and in $G^\pm_1$ as $J=1/2$. 
The ground state of $E^-$ is $J=1$ if an $A^-_2$ state is also found nearby; or better yet,  if they coincide in the infinite volume limit.
The interpretation of $G^\pm_2$ as $J = 3/2$ alone is subject to whether $J = 5/2$ or higher have significant contributions.  $J=2$ can be resolved by $B^+_1$ or $B^+_2$ or both; $J=3$  by $B^-_1$ or $B^-_2$ or both.
$J=4$ is accessible by $A^+_2$ and $J=5$ is accessible by $A^-_1$. 
There is no clean way to resolve higher spins states $J=5/2$ and $J=7/2$.

\begin{table}
\begin{tabular}{c c | l l }\toprule
$J$ & $^2D_{4h}$  &  $^2D_{4h}\;\;$ &  $J$\\
\hline
0 & $A^+_1$                                                                                           & $A^+_1\;\;$  & 0,  2, 4(2),    $\cdots$   \\
1 & $A^-_2 \oplus E^-$                                                                           &  $A^-_1\;\;$  &  5,  7,  9(2), $\cdots$ \\
2 & $A^+_1 \oplus B^+_1\oplus B^+_2 \oplus E^+$                               & $A^-_2\;\;$  & 1,  3, 5(2),  $\cdots$   \\
3 & $A^-_2 \oplus B^-_1\oplus B^-_2 \oplus 2E^-$                                 & $A^+_2\;\;$  &  4, 6, 8(2),  $\cdots$   \\
4 & $2A^+_1 \oplus A^+_2\oplus B^+_1 \oplus B^+_2 \oplus 2E^+$      &   $B^+_1\;\;$  & 2,  4, 6(2), $\cdots$     \\
5 & $ A^-_1 \oplus 2A^-_2\oplus B^-_1 \oplus B^-_2 \oplus 3E^-$         &   $B^-_1\;\;$  & 3,  5, 7(2), $\cdots$  \\
6 & $2A^+_1 \oplus  A^+_2 \oplus 2B^+_1\oplus 2B^+_2 \oplus 3E^+$ &   $B^+_2\;\;$  & 2, 4, 6(2), $\cdots$ \\
   &                                                                                                            & $B^-_2\;\;$  & 3, 5, 7(2), $\cdots$  \\
$\cdots$ & $\cdots$                                                                                 & $E^-\;\;$   & 1,  3(2), 5(3), $\cdots$   \\
   &                                                                                                           &  $E^+\;\;$    &  2,  4(2), 6(3), $\cdots$ \\
   &                                                                                                            &   & \\
1/2 & $G^\pm_1$                                                                                      & $G^\pm_1\;\;$  & 1/2,  3/2, 5/2,  $\cdots$   \\
3/2 & $G^\pm_1\oplus G^\pm_2 $                                                            & $G^\pm_2\;\;$  & 3/2, 5/2(2), 7/2(2),  $\cdots$  \\
5/2 & $ G^\pm_1\oplus 2G^\pm_2 $    & &  \\
7/2 & $2G^\pm_1\oplus 2G^\pm_2 $   & & \\
$\cdots$ & $\cdots$  \\
\bottomrule
\end{tabular}
\caption{Decomposition of angular momentum in the elongated box according to the irreps of the $^2D_{4h}$ group. 
Both the original reduction (left) and its inverse (right) are shown. Parity is indicated by the plus (even) or minus (odd) sign.}
\label{tab:2D4h}
\end{table}
%

\subsection{Phaseshift formulas in the elongated box}
\label{sec:phase_elongated}

The case for mesons has been considered in Ref.~\cite{Feng:2004ua}. 
Here we extend that approach to baryons using our unified treatment of single and double groups in Appendix~\ref{app:elongated}.
Our starting point is the real part of Eq.~\ref{eq:phaselat}, 
expressed for a given total angular momentum $J$ and partial-wave $l$,
\beq
\det  [ \mathcal{M}_{Jl M,J'l' M' }- \delta_{JJ'}  \delta_{ll'}  \delta_{MM'} \cot \delta_{Jl} ] 
=0\,.
\label{eq:phaselat2}
\eeq
The matrix $\mathcal{M}$ is adapted from 
the original one by L\"{u}scher for integer angular momentum, cubic boxes, 
and equal masses to the current case of 
half-integer angular momentum, elongated boxes (limited to $z$-direction), and unequal masses. The projection to half-integer angular momentum is achieved by a straightforward change of basis by coupling to spin-1/2,
\beq
 \mathcal{M}_{Jl M,J'l' M' } = 
\sum_{m m'm_s m_s'} \big\langle l m, {1\over 2} m_s | JM \big\rangle \big\langle l' m', {1\over 2} m'_s | J' M' \big \rangle 
 \mathcal{M}_{l m,l' m' } ,
 \label{eq:mproj}
\eeq
using Clebsch-Gordan coefficients. The modified matrix for $z$-elongated 
box (of elongation $\eta$) is 
\beq
\mathcal{M}_{l m,l' m' } (q,{\eta})= \sum_{j=[l-l'|}^{l+l'} \sum_{s=-j}^{j} 
{  (-1)^l  i^{l+l'} \over \pi^{3/2} \eta q^{j+1} } \,  Z_{js}(1,q^2,\eta)
 \times \langle l0j0|l'0\rangle  \langle lmjs|l'm'\rangle  \sqrt{{(2l+1)(2j+1)\over (2l'+1)}}.
\label{eq:mmat}
\eeq
It is customary to introduce the short-hand function for the zeta function,
\beq
\mw_{lm}(1,q^2;\eta) = \frac{\mathcal{Z}_{lm}(1,q^2;\eta)}{\pi^{3/2}\eta q^{l+1}},
\label{eq:wfun}
\eeq
so the simplest phaseshift formula reads $\cot\delta=\mw_{00}$. 
The $\mathcal{M}$ matrix is a linear combination of $\mw$ functions with purely numerical coefficients.
The dimensionless momentum $\bm q$ is defined in terms of the minimal momentum in a periodic box
of size $L$, $\bm k=(2\pi/L)\bm q$.
The generalized zeta function for $z$-elongated boxes is
\beq
\mathcal{Z}_{lm} (s,q^2;\eta) = \sum_{\bm n\in Z^3} \frac{\mathcal{Y}_{lm}(\widetilde{\bm n})}{(\widetilde{\bm n}^2-q^2)^s},
\label{eq:zfun}
\eeq
where $\mathcal{Y}_{lm}(\bm r)\equiv r^l Y_{lm}(\theta,\phi)$ are homogenous harmonic polynomials 
and the modified index 
$\widetilde{\bm n}$ is related to the cubic index $\bm n=(n_x,n_y,n_z)$ by
\beq
\widetilde{\bm n} = (n_x,n_y,n_z/\eta). 
\label{eq:psum}
\eeq
Details on how to numerically evaluate the function with elongation can be found in Refs.~\cite{Feng:2004ua,Guo:2016zos}.

The spin-projected matrix  $\mathcal{M}_{Jl M,J'l' M' }$ is still expressed in terms of angular momentum labels $Jl M$. 
Our  goal is to reduce the matrix to the irreps of the  $^2D_{4h}$ group in elongated boxes. 
Operationally, it is equivalent to the reduction of the matrix into its block diagonal form with each block having the dimension of an 
irrep.  This is achieved by another change of basis, using the basis vectors derived in Table~\ref{tab:basis2D4} in Appendix~\ref{app:elongated} where the notations used below are fully explained.
In the new basis,  $\mathcal{M}$ is block-diagonalized by irreps
\beq
 \langle \Gamma \alpha J l n |  \mathcal{M} | \Gamma' \alpha' J' l' n'  \rangle \\  =
 \sum_{MM'}  \left (C^{\Gamma \alpha n}_{Jl M}\right )^*  C^{\Gamma' \alpha' n'}_{J'l' M'}  \mathcal{M}_{Jl M,J'l'M' } 
 = \delta_{\Gamma \Gamma'} \delta_{\alpha \alpha'}   \mathcal{M}^\Gamma_{Jl n,J'l' n' },
 \eeq
where Schur's lemma in linear algebra was used in the second step. For multi-dimensional irreps, 
the matrix is diagonal in $\alpha$ and the quantization condition does not depend on it.
The final form for the phaseshift reduction is
\beq
\prod_\Gamma  \det  \left[ \mathcal{M}^\Gamma_{Jl n,J'l' n' }- \delta_{JJ'}  \delta_{ll'}  \delta_{nn'} \cot \delta_{Jl} \right ]=0.
\label{eq:phaselat3}
\eeq
If there is no multiplicity, the labels $n$ and $n'$ can be dropped.

Our results for non-zero matrix elements are given in Table~\ref{tab:me2D4h1} for integral angular momentum up to $J=4$. 
We have  exploited symmetry properties to simplify the matrix.
First, the matrix is hermitian (or symmetric if the elements are real-valued), $ \mathcal{M}^\Gamma_{J n , J' n'} = \left( \mathcal{M}^\Gamma \right)^*_{J' n', J n}$, 
so we only need to display half of the off-diagonal elements. 
Second, a lot of the $\mw$ functions vanish, or satisfy certain constraints, 
which can be traced back to how the zeta function (more specifically the spherical harmonics)
behave under the symmetry operations in the elongated box.
The properties are as follows~\cite{Leskovec:2012gb,Gockeler:2012yj}.
\begin{enumerate}[(i)]
\item 
The standard property  $Y_{l-m}=(-1)^m Y^*_{lm}$ translates directly to 
\beq \mw_{l-m}=(-1)^m \mw^*_{lm}. \label{eq:w1} \eeq  
\item 
The system is invariant under a mirror reflection about the $xy$-plane  (see Fig.~\ref{fig:box_elongated}).
It leads to $Y_{lm}(\theta,\phi)=Y_{lm}(\pi-\theta,\phi)=(-1)^{(l-m)} Y_{lm}(\theta,\phi)$, which means 
\beq \mw_{lm}=0 \text{  for  } l-m = \text{odd.  In particular  } \mw_{l0}=0  \text{  for  } l = 1,3,5,\cdots.  \label{eq:w2}  \eeq
It is valid for all systems with inversion symmetry, which leads 
to a separation into sectors by parity in the table.
\item 
The system is invariant under a $\pi/2$ rotation about the $z$-axis (or the $C_{4z}$ element of $^2D_{4h}$). 
It leads to the constraint $e^{im\pi/2}=1$ due to the $e^{im\phi}$ dependence in $Y_{lm}$.
This means 
\beq \mw_{lm}=0 \text{  for  }  m \neq  0, 4, 8, \cdots, \text{ regardless of } l.  \label{eq:w3}  \eeq 
\item 
The system is invariant under a mirror reflection about the $xz$-plane,
which leads to 
\beq Y_{lm}(\theta,\phi)=Y_{lm}(\theta,2\pi-\phi)= Y^*_{lm}(\theta,\phi)\,.  
\label{eq:w4}  
\eeq 
This means all the zeta functions are real in the elongated box. 
\item 
Furthermore, combining conditions (i), (iii), (iv) yields a new condition  
\beq \mw_{l-m}= \mw_{lm} \text{  for  }  m=  0, 4, 8, \cdots, \label{eq:w5} \eeq
that is, there is  no difference between $m$ and $-m$ for the allowed values of $m$ in the elongated box.
\end{enumerate}
Our results agree with those in Ref.~\cite{Feng:2004ua}. 
For half-integral angular momentum up to $J=7/2$ our results for the two irreps $G_1$ and $G_2$ are new 
and are  given in Table~\ref{tab:me2D4h2}. There is two-fold multiplicity for $J=7/2$ so the $n$ and $n'$ labels are kept explicit.



%
\begin{table*}[p]
\begin{tabular}{c}
$      
\renewcommand{\arraystretch}{2.0}
\begin{array}{c| c c | c c | c }
\toprule
\Gamma & J  & n  &  J'    & n'  & \mathcal{M}^\Gamma_{J n , J' n'}   \\
\hline 
A^+_1 & 0 & 1  & 0 & 1  & \mw_{00}  \\
    & 0 & 1  & 2 & 1  & -\mw_{20} \\
    & 0 & 1  & 4 & 1  & \mw_{40} \\
    & 0 & 1  & 4 & 2  & \sqrt{2} \mw_{44}  \\
    & 2 & 1  & 2 & 1  & \mw_{00} + {2\sqrt{5}\over 7} \mw_{20} + {6\over 7} \mw_{40} \\
    & 2 & 1  & 4 & 1  & - {6\over 7} \mw_{20} - {20\sqrt{5}\over 77}\mw_{40} - {15\over 11} \sqrt{{5\over 13}}\mw_{60}  \\
    & 2 & 1  & 4 & 2  & \frac{4\sqrt{10}}{11}  \mw_{44}-\frac{15}{11} \sqrt{\frac{2}{13}} \mw_{64} \\
    
    & 4 & 1  & 4 & 1  & \mw_{00} + {20\sqrt{5}\over 77}\mw_{20} +  {486\over 1001} \mw_{40}  
        +  {20 \over 11 \sqrt{13}} \mw_{60}  + {490 \over 143 \sqrt{13}} \mw_{80}  \\
        
    & 4 & 1  & 4 & 2  & \frac{54 \sqrt{2}}{143} \mw_{44}-\frac{12}{11} \sqrt{\frac{10}{13}} \mw_{64}+\frac{21}{13} \sqrt{\frac{10}{187}} \mw_{84}  \\
    
    & 4 & 2  & 4 & 2  & \mw_{00}-\frac{4\sqrt{5} }{11} \mw_{20}+\frac{54}{143} \mw_{40}-\frac{4 }{11 \sqrt{13}}\mw_{60}+\frac{7 }{143 \sqrt{17}} \mw_{80}+21 \sqrt{\frac{10}{2431}} \mw_{88} \\
    
\hline 
A^-_2 & 1 & 1  & 1 & 1  & \mw_{00} + {2\over \sqrt{5}} \mw_{20}         \\
    & 1 & 1  & 3 & 1  & - 3\sqrt{{3\over 35}}\mw_{20} - {4\over \sqrt{21}} \mw_{40}     \\
    & 3 & 1  & 3 & 1  & \mw_{00} + {4\over 3\sqrt{5}} \mw_{20} + {6\over 11} \mw_{40} + {100\over 33\sqrt{13}} \mw_{60} \\
    \hline
A^+_2    & 4 & 1  & 4 & 1  & \mw_{00}-\frac{4\sqrt{5} }{11} \mw_{20}+\frac{54}{143} \mw_{40}-\frac{4 }{11 \sqrt{13}} \mw_{60}+\frac{7 }{143 \sqrt{17}} \mw_{80} -21 \sqrt{\frac{10}{2431}} \mw_{88} \\
    
\hline 
B^+_{1/2} & 2 & 1  & 2 & 1  & \mw_{00}-\frac{2 \sqrt{5} }{7}\mw_{20}+\frac{1}{7} \mw_{40} \pm \sqrt{\frac{10}{7}} \mw_{44} \\

 & 2 & 1  & 4 & 1  & -\frac{\sqrt{15} }{7} \mw_{20}+\frac{30\sqrt{3}}{77}  \mw_{40}-\frac{5}{11} \sqrt{\frac{3}{13}} \mw_{60} \pm \frac{2}{11} \sqrt{\frac{30}{7}} \mw_{44} \mp \frac{5}{11} \sqrt{\frac{42}{13}} \mw_{64}  \\  
    & 4 & 1  & 4 & 1  &\mw_{00}+\frac{8\sqrt{5} }{77} \mw_{20}-\frac{27}{91} \mw_{40}-\frac{2 }{\sqrt{13}} \mw_{60}+\frac{196 }{143 \sqrt{17}} \mw_{80} \\
    &&&&& \pm\frac{81}{143} \sqrt{\frac{10}{7}} \mw_{44}\pm\frac{6}{11} \sqrt{\frac{14}{13}} \mw_{64}\pm\frac{42}{13} \sqrt{\frac{14}{187}} \mw_{84} \\
\hline
B^-_{1/2}    & 3 & 1  & 3 & 1  &\mw_{00}-\frac{7}{11} \mw_{40}+\frac{10 }{11 \sqrt{13}}\mw_{60}\mp\frac{\sqrt{70}}{11}  \mw_{44}\mp\frac{10}{11} \sqrt{\frac{14}{13}} \mw_{64}  \\

\hline 
E^- & 1 & 1 &  1 & 1 &   \mw_{00} - {1\over \sqrt{5}} \mw_{20}   \\
   & 1 & 1 &  3 & 1 &  -3 \sqrt{\frac{2}{35}} \mw_{20}+\sqrt{\frac{2}{7}} \mw_{40}  \\
   & 3 & 1 &  3 & 1 & \mw_{00}+\frac{1}{\sqrt{5}} \mw_{20}+\frac{1}{11} \mw_{40}-\frac{25 }{11 \sqrt{13}}  \mw_{60} \\
   & 3 & 2 &  3 & 2 & \mw_{00}-\frac{ \sqrt{5}}{3} \mw_{20}+\frac{3}{11} \mw_{40}-\frac{5 }{33 \sqrt{13}} \mw_{60}\\
\hline
E^+   & 2 & 1 &  2 & 1 &  \mw_{00}+\frac{\sqrt{5} }{7} \mw_{20}-\frac{4}{7} \mw_{40} \\                                
   & 2 & 1 &  4 & 1 & -\frac{\sqrt{30}}{7}  \mw_{20}-\frac{5\sqrt{6}}{77}  \mw_{40}+\frac{10}{11} \sqrt{\frac{6}{13}} \mw_{60} \\      
   & 4 & 1 &  4 & 1 & \mw_{00}+\frac{17 \sqrt{5}}{77} \mw_{20}+\frac{243 }{1001}\mw_{40} -\frac{1}{11 \sqrt{13}} \mw_{60} -\frac{392 }{143 \sqrt{17}} \mw_{80} \\
   & 4 & 2 &  4 & 2 & \mw_{00}-\frac{4\sqrt{5} }{11} \mw_{20}+\frac{54}{143} \mw_{40}-\frac{4 }{11 \sqrt{13}}\mw_{60}+\frac{7 }{143 \sqrt{17}} \mw_{80}\\
\bottomrule
\end{array}
$      
\end{tabular}
\caption{Non-zero reduced matrix elements in the elongated box ($D_{4h}$ symmetry group)
for integral angular momentum up to $J=4$.  
The $B_{1}$ and $B_2$ irreps are combined as indicated by the upper/lower signs.
The matrix is symmetric in $J$ and $J'$ in each irrep-parity sector. 
The results are separated according to parity $(-1)^l$.
}
\label{tab:me2D4h1}
\end{table*}
\begin{table}
\begin{tabular}{c}
$      
\renewcommand{\arraystretch}{1.8}
\begin{array}{c | c c | c c | c }
\toprule
\Gamma  & J           & n & J'          & n' & \mathcal{M}^\Gamma_{J l n , J' l' n' }                                                            \\
\hline  
G^\pm_1 & {1\over 2}  & 1 & {1\over 2}  & 1  & \mw_{00}                                                                                          \\
        & {1\over 2}  & 1 & {3\over 2}  & 1  & \pm \sqrt{\frac{2}{5}} \mw_{20}                                                                     \\
        & {1\over 2}  & 1 & {5\over 2}  & 1  & -\sqrt{\frac{3}{5}} \mw_{20}                                                                        \\
        & {1\over 2}  & 1 & {7\over 2}  & 1  & -\frac{2}{3} \mw_{40}                                                                               \\
        & {1\over 2}  & 1 & {7\over 2}  & 2  & \mp \frac{2}{3} \sqrt{2} \mw_{44}                                                         \\
  
        & {3\over 2}  & 1 & {3\over 2}  & 1  & \mw_{00}+\frac{1}{\sqrt{5}} \mw_{20}                                                        \\
        & {3\over 2}  & 1 & {5\over 2}  & 1  & \mp \frac{1}{7} \sqrt{\frac{6}{5}} \mw_{20} \mp \frac{2\sqrt{6}}{7}  \mw_{40}                         \\
        & {3\over 2}  & 1 & {7\over 2}  & 1  &-\frac{9}{7} \sqrt{\frac{2}{5}} \mw_{20}-\frac{5\sqrt{2} }{21} \mw_{40}                                \\
        & {3\over 2}  & 1 & {7\over 2}  & 2  & \frac{2}{3} \mw_{44}                                                                      \\
  
        & {5\over 2}  & 1 & {5\over 2}  & 1  & \mw_{00}+\frac{8 }{7 \sqrt{5}}\mw_{20}+\frac{2}{7} \mw_{40}                                             \\
        & {5\over 2}  & 1 & {7\over 2}  & 1  & \pm \frac{2 }{7 \sqrt{15}}\mw_{20}\pm\frac{10\sqrt{3}}{77}  \mw_{40}\pm\frac{50 }{11 \sqrt{39}} \mw_{60} \\
        & {5\over 2}  & 1 & {7\over 2}  & 2  &\mp\frac{2\sqrt{6}}{11}  \mw_{44}\pm\frac{10}{11} \sqrt{\frac{10}{39}} \mw_{64}    \\
  
        & {7\over 2}  & 1 & {7\over 2}  & 1  & \mw_{00}+\frac{5}{21} \sqrt{5} \mw_{20}+\frac{27}{77} \mw_{40}+\frac{25 \mw_{60}}{33 \sqrt{13}}           \\
        & {7\over 2}  & 1 & {7\over 2}  & 2  & \frac{3\sqrt{2}}{11}  \mw_{44}-\frac{5}{11} \sqrt{\frac{10}{13}} \mw_{64}         \\
        & {7\over 2}  & 2 & {7\over 2}  & 2  & \mw_{00}-\frac{\sqrt{5}}{3}  \mw_{20}+\frac{3}{11} \mw_{40}-\frac{5}{33 \sqrt{13}}                \mw_{60} \\
\hline  
G^\pm_2        & {3\over 2}  & 1 & {3\over 2}  & 1  & \mw_{00}-\frac{1}{\sqrt{5}} \mw_{20}          \\
        & {3\over 2}  & 1 & {5\over 2}  & 1  & \pm \frac{6 }{7 \sqrt{5}} \mw_{20} \mp \frac{2}{7} \mw_{40}       \\
        & {3\over 2}  & 1 & {5\over 2}  & 2  & \mp 2 \sqrt{\frac{2}{7}} \mw_{44}    \\
        & {3\over 2}  & 1 & {7\over 2}  & 1  & -\frac{3\sqrt{2} }{7} \mw_{20}+\frac{ \sqrt{10}}{7} \mw_{40}     \\
        & {3\over 2}  & 1 & {7\over 2}  & 2  & \frac{2 }{\sqrt{21}}  \mw_{44}  \\
 
        & {5\over 2}  & 1 & {5\over 2}  & 1  &\mw_{00}+\frac{2 }{7 \sqrt{5}} \mw_{20} -\frac{3}{7} \mw_{40} \\
        & {5\over 2}  & 1 & {5\over 2}  & 2  & \sqrt{\frac{2}{7}} \mw_{44}    \\        
        & {5\over 2}  & 2 & {5\over 2}  & 2  & \mw_{00}-\frac{2\sqrt{5} }{7} \mw_{20}+\frac{1}{7} \mw_{40}   \\

        & {5\over 2}  & 1 & {7\over 2}  & 1  & \pm \frac{\sqrt{2}}{7}  \mw_{20}\pm\frac{8\sqrt{10} }{77} \mw_{40}\mp\frac{5}{11} \sqrt{\frac{10}{13}} \mw_{60} \\
        & {5\over 2}  & 1 & {7\over 2}  & 2  & \mp \frac{8}{11} \sqrt{\frac{3}{7}} \mw_{44} \mp \frac{10}{11} \sqrt{\frac{35}{39}} \mw_{64}   \\
        & {5\over 2}  & 2 & {7\over 2}  & 1  & \pm \frac{4}{11} \sqrt{\frac{5}{7}} \mw_{44} \mp \frac{10}{11} \sqrt{\frac{7}{13}} \mw_{64}  \\
        & {5\over 2}  & 2 & {7\over 2}  & 2  & \mp\frac{1}{7} \sqrt{\frac{10}{3}} \mw_{20}\pm\frac{10}{77} \sqrt{6} \mw_{40}\mp\frac{5}{11} \sqrt{\frac{2}{39}} \mw_{60} \\        
  
        & {7\over 2}  & 1 & {7\over 2}  & 1  & \mw_{00}+\frac{\sqrt{5} }{7} \mw_{20}-\frac{9}{77} \mw_{40}-\frac{15 }{11 \sqrt{13}}     \mw_{60}\\
        & {7\over 2}  & 1 & {7\over 2}  & 2  & \frac{3}{11} \sqrt{\frac{30}{7}} \mw_{44}+\frac{5}{11} \sqrt{\frac{14}{39}} \mw_{64}  \\
        & {7\over 2}  & 2 & {7\over 2}  & 2  & \mw_{00}-\frac{ \sqrt{5}}{21} \mw_{20}-\frac{39}{77} \mw_{40}+\frac{25 }{33 \sqrt{13}}  \mw_{60} \\

\bottomrule
\end{array}
$          
\end{tabular}
\caption{Non-zero reduced matrix elements in the elongated box ($^2D_{4h}$ symmetry group)
 for half-integral angular momentum up to $J=7/2$.  The even/odd parity sectors are indicated by the upper/lower signs.
The matrix is symmetric in $Jn$ and $J'n'$ in each irrep-parity sector.}
\label{tab:me2D4h2}
\end{table}
%


The final step is to determine the phaseshift using the matrix elements.
The $A^+_1$ sector couples to $J=0, 2, 4$ with two-fold multiplicity in $J=4$. 
The full treatment will lead to a $4\times 4$ matrix in  $ \mm_{J  n , J' n' } $. If the mixing with $J=4$ can be ignored, we expect 
only a  $2\times 2$ matrix $\mm_{JJ'}$ (the multiplicity label is suppressed) and  Eq.~\ref{eq:phaselat3} takes the simple form,
\beq
\left |
\begin{array}{cc}       
\mm_{00} - \cot\delta_0 & \mm_{02}     \\
\mm_{20}                & \mm_{22} - \cot\delta_2    \\
\end{array}             
\right |                
=0.                     
\label{eq:detA1p}
\eeq
The solution is
\beq
 {\bm A^+_1 \;\;\text{sector}:}  \;\;\;
\cot\delta_0 
= \mw_{00} +  {\mw^2_{20} \over \cot\delta_2 - (\mw_{00} + {2\sqrt{5}\over 7} \mw_{20} + {6\over 7} \mw_{40})}  .
\eeq
So the determination of $\delta_0$ generally involves $ \mw_{00}$, $ \mw_{20}$,  and $\mw_{40}$, as well as knowledge of the $\delta_2$.
If the coupling to $J=2$ and higher can be ignored, one gets the simple formula for the $J=0$ phaseshift $\cot\delta_0 = \mw_{00}$. 
This is the only irrep that has access to the $J=0$ resonance  in the elongated box.

The $A^-_2$ sector at the given cutoff couples to $J=1, 3$ with no multiplicities. 
The $\delta_1$ can be determined via
\beq
\cot\delta_1 = \mm_{11} + {\mm_{13}  \mm_{31} \over \cot\delta_3 - \mm_{33} }
\eeq
which involves $ \mw_{l0}$ with $l=0,2,4,6$ and $\delta_3$.
If the coupling to $J=3$ and higher can be ignored, one gets the simple formula for the $J=1$  
phaseshift 
\beq
 {\bm A^-_2 \;\;\text{sector}:}  \;\;\;
\cot\delta_1 = \mw_{00} + {2\over \sqrt{5}} \mw_{20} 
\eeq
which has been used in Ref.~\cite{Guo:2016zos,Pelissier:2012pi}.

As an example of how to treat multiplicity, we write down the full $3\times 3$ matrix equation 
in the $E^-$  sector 
\beq                       
\left |                    
\begin{array}{cccc}        
\mm_{11,11} - \cot\delta_1 & \mm_{11,31}                 &  0               \\
\mm_{31,11}                &  \mm_{31,31} - \cot\delta_{3} &0        \\
0            & 0            & \mm_{32,32} - \cot\delta_{3}    \\
\end{array}                
\right |                   
=0.                        
\label{eq:detE}            
\eeq                       
It yields the solution 
\beq
\cot\delta_1 = \mm_{11,11} + {\mm_{11,31}  \mm_{31,11} \over \cot\delta_3 - \mm_{31,31} }
;\;\;\; \cot\delta_3 =  \mm_{32,32}. 
\eeq
Furthermore, if the $J=3$ state can be ignored,  we get the simple formula
\beq
{\bm E^- \;\;\text{sector}:}  \;\;\;  \cot\delta_1 =  \mw_{00} - {1\over \sqrt{5}} \mw_{20}.
\eeq
Because of zero coupling between $Jn$ combinations 31 and 32, we get clean access to $\delta_3$ in this sector up to the cutoff $J=4$,
\beq
{\bm E^- \;\;\text{sector}:}  \;\;\;  \cot\delta_3 =   \mw_{00}-\frac{ \sqrt{5}}{3} \mw_{20}+\frac{3}{11} \mw_{40}-\frac{5 }{33 \sqrt{13}} \mw_{60}.
\eeq

In similar fashion, the $J=2$ channel can be accessed though the $B^+_{1/2}$ and $E^+$ sectors. 
The $J=3$ channel can also be accessed though the $B^-_{1/2}$ sectors.
The best sector for $J=4$ is $A^+_{2}$ where $J=4$ is the ground state.

For half-integral angular momentum, the $G^\pm_1$ sector gives the only access to spin-1/2 phaseshifts $\delta_{{1\over 2} 0}$ (for $S_{11}$)  
and $\delta_{{1\over 2} 1}$ (for $P_{11}$), assuming a spin-0 meson. 
However, they mix with $J=3/2, 5/2, 7/2$ with two-fold multiplicity in $J=7/2$. 
The full mixing with cutoff at $J=7/2$ entails a $5\times 5$ matrix in $\mm_{J n , J' n' } $. 
The mixing of $J=1/2$ with $J=3/2$ and $J=5/2$ involves $\mm_{20}$; while $J=1/2$ and $J=7/2$ mixing involves $\mm_{40}$ and $\mm_{44}$.
If we assume coupling to $J=7/2$ and higher can be ignored, then $\delta_{1\over2}$ can be determined via the relation
\beq                                                     
\left |                                     
\begin{array}{ccc}                                     
\mm_{{1\over 2}{1\over 2}} - \cot\delta_{{1\over 2}} & \mm_{{1\over 2}{3\over 2}}                           & \mm_{{1\over 2}{5\over 2}} \\
\mm_{{3\over 2}{1\over 2}}                           & \mm_{{3\over 2}{3\over 2}} - \cot\delta_{{3\over 2}} & \mm_{{3\over 2}{5\over 2}} \\
\mm_{{5\over 2}{1\over 2}}                           & \mm_{{5\over 2}{3\over 2}}                           & \mm_{{5\over 2}{5\over 2}} - \cot\delta_{{5\over 2}}
\end{array}                                              
\right |                                           
=0,                                                      
\label{eq:detG1} 
\eeq
where the multiplicity and parity labels are suppressed. 
The determinant in Eq.~\ref{eq:detG1} involves only the product of the three off-diagonal elements and their squares. 
Two of them ($ \mm_{{1\over 2}{3\over 2}} $ and $ \mm_{{3\over 2}{5\over 2}} $) differ by a sign for even/odd parity, 
and one the same sign ($ \mm_{{1\over 2}{5\over 2}} $). This means that Eq.~\ref{eq:detG1}  is independent of parity; or
 $\delta_{{1\over 2} 0}$ and $\delta_{{1\over 2} 1}$ obey the same phaseshift formula, so do 
$\delta_{{3\over 2} 1}$ and $\delta_{{3\over 2} 2}$, and $\delta_{{5\over 2} 2}$ and $\delta_{{5\over 2} 3}$. 
So we can suppress the partial-wave $l$ label in $\delta_{Jl}$.
In fact, the same conclusion extends to the entire $G^\pm_1$ sector. 
If mixing with only $J=3/2$ is considered, we have
\beq
 {\bm G^\pm_1 \;\;\text{sector}:}  \;\;\;
\cot\delta_{1\over 2} 
= \mw_{00} 
+  { \frac{2}{5} \mw^2_{20} \over \cot\delta_{3\over 2} - (\mw_{00}+\frac{1}{\sqrt{5}} \mw_{20})  }. 
\label{eq:phG1}
\eeq
The determination of spin-1/2 resonances requires $\mw_{00}$ and $\mw_{20}$ and $\delta_{3\over 2}$.
Only when coupling with $J=3/2$ can be ignored can one obtain the simplest formula for the Roper ($P11$) and $S_{11}$ phaseshifts 
$\cot\delta_{1\over 2} = \mw_{00}.$ 
On the other hand, if $\delta_{1\over 2}$ has been independently determined, Eq.\ref{eq:phG1} can be used to access $\delta_{3\over 2}$.

In the $G^\pm_2$ sector, the leading contribution is $J=3/2$, followed by $J=5/2$ and $J=7/2$ which both have two-fold multiplicity.  
The full mixing up to $J=7/2$ also entails a $5\times 5$ matrix in $\mm_{J n , J' n' } $. 
If we ignore mixing with $J=7/2$, the phaseshift relation is given by the $3\times 3$ matrix equation,
\beq                                                     
\left |                                     
\begin{array}{ccc}                                     
\mm_{{3\over 2}1,{3\over 2}1} - \cot\delta_{{3\over 2}} & \mm_{{3\over 2}1,{5\over 2}1}                           & \mm_{{3\over 2}1,{5\over 2}2} \\
\mm_{{5\over 2}2,{3\over 2}1}                           & \mm_{{5\over 2}1,{5\over 2}1} - \cot\delta_{{5\over 2}} & 0 \\
\mm_{{5\over 2}2,{3\over 2}1}                           & 0     & \mm_{{5\over 2}2,{5\over 2}2} - \cot\delta_{{5\over 2}}
\end{array}                                              
\right |                                           
=0,                                                      
\label{eq:detG2}                                                                  
\eeq
which has no coupling between the two multiplicities of $J=5/2$.
The solution is
\beq
\cot\delta_{{3\over 2}} =\mm_{{3\over 2}1,{3\over 2}1} 
+ { \mm_{{3\over 2}1,{5\over 2}1}   \mm_{{3\over 2}1,{5\over 2}2}   \over \cot\delta_{{5\over 2}}  - \mm_{{5\over 2}1,{5\over 2}1} }
+ { \mm^2_{{5\over 2}2,{5\over 2}2}    \over \cot\delta_{{5\over 2}}  - \mm_{{5\over 2}2,{5\over 2}2} }.
\label{eq:phG2}
\eeq
If $J=5/2$ can be ignored, one gets the simple phaseshift  formula
 \beq
  {\bm G^\pm_2 \;\;\text{sector}:}  \;\;\;
 \cot\delta_{{3\over 2}} = \mw_{00} - {1 \over \sqrt{5}} \mw_{20}.  
 \eeq 
 This gives the best access to the $\Delta$ resonance in the elongated box.
  On the other hand, $\delta_{{5\over 2}}$  can be extracted in this sector  if $\delta_{{3\over 2}}$ has been independently determined.

\section{Phaseshift reduction in the cubic box }

In this section,  we revisit the cubic case using the same 
approach developed in the elongated case. We find such an exercise instructive 
in at least  a couple of ways. 
First, it can be used to perform consistency checks and validation on the elongated results by comparing with known results.
Second, it can serve as a basis for exploring the relationship between the two cases, 
providing valuable insight into how the results transition from one to the other.

Going from elongated to cubic entails an increase in symmetry. 
The situation is depicted in Fig.~\ref{fig:box_cubic}.
The basic symmetry group is called the octahedral (or cubic) group $O$ which has 24 elements and 5 irreps.
The $O$ group is another finite subgroup of the continuum rotation group $SO(3)$.
Although the $O$ group is sufficient in describing integral angular momentum in the cubic box, 
its double-covered group $^2O$ is needed for half-integral angular momentum, which has 48 elements and 8 irreps. 
The full symmetry group in the cubic box must also include space inversion (parity), 
denoted by $^2O_{h}$, which has 96 elements and 16 irreps.  
The full technical details of the $^2O_{h}$ group are given in Appendix~\ref{app:cubic}.
\begin{figure}
\includegraphics[scale=0.7]{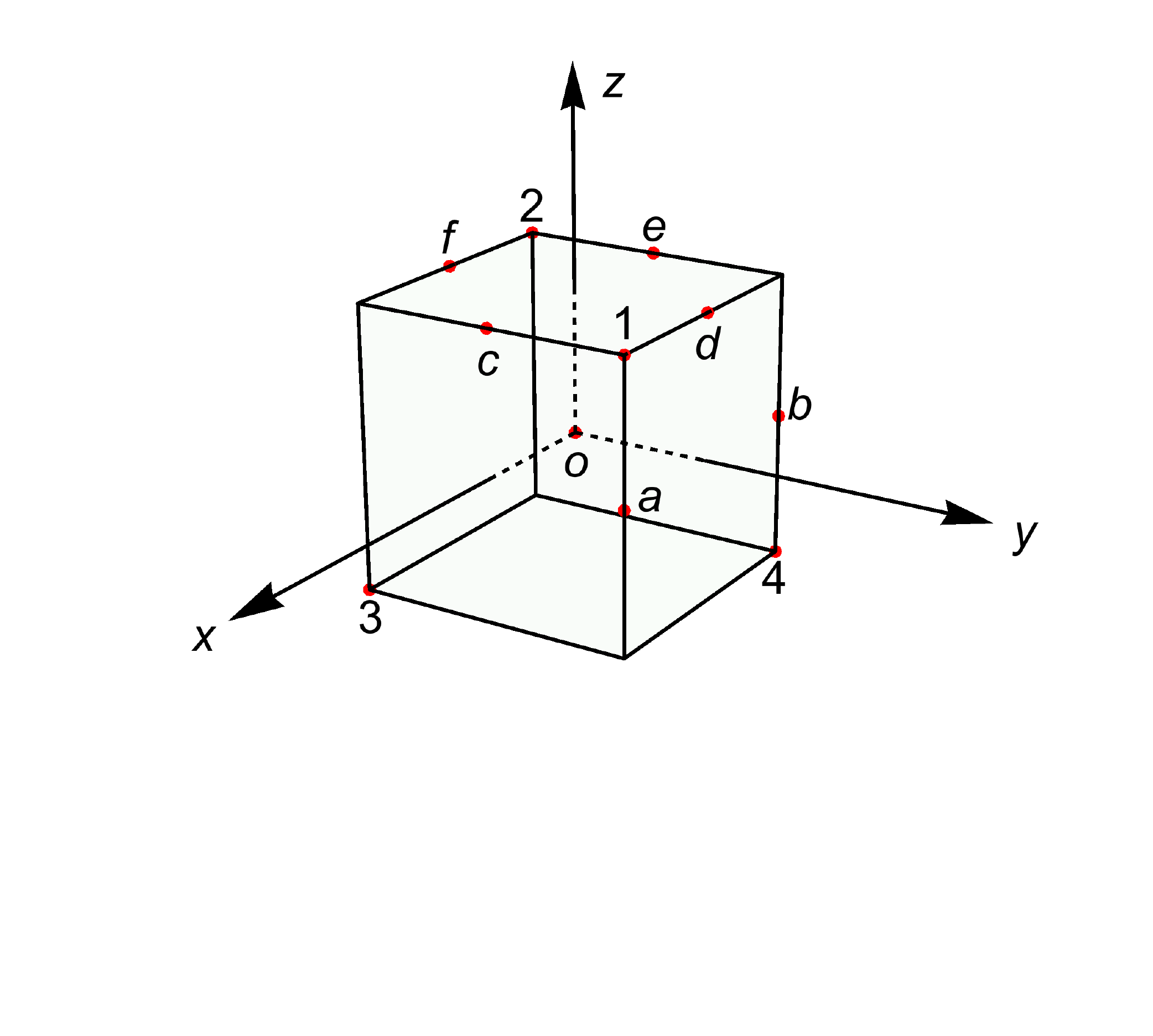}
\vspace{-2.5cm}
\caption{The 24 symmetry operations in the cubic box that form the octahedral group $O$.
They are divided into 5 conjugacy classes: 
the identity ($I$);  six $\pi/2$ rotations about the 3 axes; 
three $\pi$ rotations about the 3 axes;  
eight $2\pi/3$ rotations about 4 body diagonals denoted by 1, 2, 3, 4; 
and six $\pi$ rotations about axes  parallel to 6 face diagonals denoted by $a, b, c, d, e, f$. 
The operations are performed in a right-hand way with the thumb pointing from the center to the various symmetry points. 
The full details of the group, with the inclusion of half-integer spin and spatial inversion, 
are given in Appendix~\ref{app:cubic}.}
\label{fig:box_cubic}
\end{figure}


In the cubic box,
instead of the infinite sequence of irreps for $SU(2)$ representing angular momentum (both integer and half-integer) in the infinite volume,  only 16 possibilities (the irreps of the $^2O_h$ group) exist for angular momentum 
classification of states. 
The same decomposition method as in the elongated case, but using the characters in Table~\ref{tab:char2O} in the appendix,  leads to Table~\ref{tab:2Oh} for angular momentum resolution in the cubic box.
\begin{table}
\begin{tabular}{c c | l l}\toprule
$J$ & $^2O_h$  &  $^2O_h\;\;$ &  J\\
\hline
0 & $A^+_1$                                                               & $A^+_1\;\;$  & 0, 4, 6, $\cdots$ \\
1 & $T^-_1$                                                                & $A^-_1\;\;$  & 9, 13, 15, $\cdots$ \\
2 &  $T^+_2\oplus E^+$                                             & $T^-_1\;\;$  & 1, 3, 5(2),  $\cdots$ \\
3 & $A^-_2 \oplus T^-_1\oplus T^-_2$                       &  $T^+_1\;\;$  & 4, 6, 8(2),  $\cdots$  \\
4 & $A^+_1 \oplus E^+ \oplus T^+_1 \oplus T^+_2$  & $T^+_2\;\;$  & 2, 4,  6(2), $\cdots$ \\
5 & $ E^- \oplus  2T^-_1 \oplus T^-_2$                      & $T^-_2\;\;$  & 3, 5, 7(2), $\cdots$ \\
6 & $ A^+_1 \oplus A^+_2 \oplus E^+ \oplus  T^+_1 \oplus 2T^+_2$                    & $E^+\;\;$     & 2, 4, 6, $\cdots$ \\
&                                                                                 &  $E^-\;\;$     & 5, 7, 9,  $\cdots$ \\
$\cdots$ & $\cdots$                                                    &  $A^-_2\;\;$  & 3, 7, 9,  $\cdots$  \\
&                                                                                 & $A^+_2\;\;$  &  6, 10, 12,   $\cdots$ \\
&                                                                                 & & \\

1/2 & $G^\pm_1$                                                        & $G^\pm_1\;\;$  & 1/2, 7/2, 9/2,  $\cdots$ \\
3/2 & $H^\pm$                                                            &  $H^\pm\;\;$      & 3/2, 5/2, 7/2,  $\cdots$ \\
5/2 & $G^\pm_2 \oplus H^\pm $                                 & $G^\pm_2\;\;$  & 5/2, 7/2, 11/2,  $\cdots$ \\
7/2 & $G_1^\pm\oplus G^\pm_2 \oplus H^\pm $        & &\\
$\cdots$ & $\cdots$  \\
\bottomrule
\end{tabular}
\caption{Decomposition of angular momentum in the cubic box according to the irreps of the $^2O_h$ group. 
Both the original decomposition (left) and its inverse (right) are shown. The number in parentheses indicates the multiplicity of that $J$ in that irrep. }
\label{tab:2Oh}
\end{table}
It shows that angular momenta $J= 0, 1/2, 1,3/2$ correspond to single irreps.
It is safe to identify the ground state in the $A^+_1$ irrep as $J=0$, and in $G^\pm_1$ as $J=1/2$ because the gap to the next contributing $J$ is 3 units away. The identification of the ground state in $T^-_1$ as $J=1$ would be reasonable as long as no states in the $A^-_2$ and $T^-_2$ irrep are close in energy since this would indicate that they are part of an infinite-volume $J=3$ multiplet. 
The ground state of $E^+$ is $J=2$ if a $T^+_2$ state is also found nearby.
Likewise, the interpretation of $H^\pm$ as $J = 3/2$ alone is subject to whether $G^\pm_2$ could move in to make $J = 5/2$.


Next we carry out the same phaseshift reduction procedure as in the elongated case,  
but using the basis vectors for the cubic box given in Table~\ref{tab:basis2D4} in the appendix.
The results for the matrix elements are shown in Table~\ref{tab:me2Oh1} for integral angular momentum,
and in Table~\ref{tab:me2Oh2} for half-integral angular momentum.
Since the cubic case has higher symmetry than the elongated one, there are more symmetry properties on the $\mw$ 
function (or equivalently the zeta function) that can be used to simplify the matrix.
More conditions can be found by the general transformation on the zeta function~\cite{Luscher:1990ux},
\beq
\sum_{m'=-l}^{l} D_{mm'}^{(l)}(R) \mathcal{Z}_{lm'}(s,q^2)= \mathcal{Z}_{lm}(s,q^2),
\eeq
where the Wigner-D functions can be evaluated by using the Euler angles of the cubic group rotations given in Table~\ref{tab:irrep2O}. 
In addition to the constraints in the elongated box discussed in Eq.~\ref{eq:w1} to Eq.~\ref{eq:w5}, 
we have (up to $l=8$)
\beq
\mw_{20} =0,\;
\mw_{44} = {\sqrt{70}\over 14}  \mw_{40},\;
\mw_{64} = - {\sqrt{14}\over 2}  \mw_{60}, \;
\mw_{84} =  {\sqrt{154}\over 33}  \mw_{80}, \;
\mw_{88} =  {\sqrt{1430}\over 66}  \mw_{80}.
\label{eq:wcubic}
\eeq
Our results for half-integer angular momentum in Table~\ref{tab:me2Oh2} agree with those
 in Ref.~\cite{Bernard:2008ax}, 
 after accounting for the $\sqrt{2l+1}$ factor in the definition of $\mw$ functions.

Now we turn to the phaseshift formulas. 
In the cubic box, the only access to  phaseshift  $\delta_0$  is in the 
\beq
{\bm A^+_1 \;\;\text{sector}:}  \;\;\;  \cot\delta_0 =  \mw_{00},
\eeq
if we can ignore coupling to $J=4$ and higher phaseshifts, which are expected to be small at low energies 
(when $\delta_l(k)\propto k^{2l+1}$).

The only access to phaseshift $\delta_1$  is in the 
\beq
{\bm T^-_1 \;\;\text{sector}:}  \;\;\;  \cot\delta_1 =  \mw_{00},
\eeq
if we can ignore coupling to $J=3$ and higher. 
Although the formula for $\delta_1$ is the same as that for $\delta_0$,
 differences can arising from mixing with higher states, 
and from operators constructed under the different irreps to access the states.

The  $\delta_2$ phaseshift can be accessed either in the
\beq
{\bm E^+ \;\;\text{sector}:}  \;\;\;  \cot\delta_2 =  \mw_{00}+{6\over 7} \mw_{40},
\eeq
or  in the
\beq
{\bm T^+_2 \;\;\text{sector}:}  \;\;\;  \cot\delta_2 =  \mw_{00}-{4\over 7} \mw_{40}.
\eeq
Similarly, the best access to  $\delta_3$ phaseshift is in the  $A^-_2$ sector or $T^-_2$ sector.
The best access to  $\delta_4$ phaseshift is in the $T^+_1$ sector.

For half-integer spin, the only access to spin-1/2 resonance phaseshift  is in the 
\beq
{\bm G^\pm_1 \;\;\text{sector}:}  \;\;\;  \cot\delta_{1\over 2} =  \mw_{00}.
\eeq
It has a large gap to the next state it mixes with, $J=7/2$, which involves $\mw_{40}$. 
Note that $\delta_{1\over 2}$ stands for $\delta_{{1\over 2}0}$  for  $G^+_1$,  and $\delta_{{1\over 2}1}$  for  $G^-_1$. 
For $\pi N$ scattering, they correspond to the $S_{11}$ and $P_{11}$ (Roper) phaseshifts, respectively. 
We see that obey the same formula in the cubic box.

The only access to phaseshift $\delta_{3\over 2}$ is in the $H^\pm$ sector, with leading contribution
\beq
{\bm H^\pm \;\;\text{sector}:}  \;\;\;  \cot\delta_{3\over 2} =  \mw_{00}.
\eeq
This is the best sector to extract the phaseshift of $\Delta$ resonance. 
Mixing with higher states $J=5/2$ and $J=7/2$ can be included using the provided matrix elements.

The best access to $\delta_{5\over 2}$ phaseshift is in the 
\beq
{\bm G^\pm_2 \;\;\text{sector}:}  \;\;\;  \cot\delta_{5\over 2} =   \mw_{00}-{4\over 7} \mw_{40}.
\eeq
It couples with $J=7/2$ and higher states.
The combined results from the different sectors can help isolate the phaseshifts for low-lying baryon resonances 
in $\pi N$ scattering in the cubic box.

\begin{table}
\begin{tabular}{c}
$      
\renewcommand{\arraystretch}{1.8}
\begin{array}{c | c | c | l }
\toprule
\Gamma & J  & J'  & \mathcal{M}^\Gamma_{J, J'}                                         \\
\hline 
A^+_1    & 0  & 0   & \mw_{00}                                                                 \\
       & 0  & 4   &  {36 \over \sqrt{21} } \mw_{40} \\
      & 4 & 4  & \mw_{00}+\frac{108}{143} \mw_{40}+\frac{80}{11 \sqrt{13}} \mw_{60}+\frac{560}{143 \sqrt{17}}  \mw_{80}\\
        
\hline 
A^-_2  & 3  & 3  & \mw_{00}-\frac{12}{11} \mw_{40}+\frac{80}{11 \sqrt{13}} \mw_{60} \\
     
\hline 
E^+  & 2 &  2  &  \mw_{00}+\frac{6}{7} \mw_{40} \\
  
   & 2 &  4  &  -\frac{40\sqrt{3} }{77} \mw_{40}-\frac{30}{11} \sqrt{\frac{3}{13}} \mw_{60} \\
   & 4  &  4  &  \mw_{00}+\frac{108}{1001} \mw_{40}-\frac{64}{11 \sqrt{13}}  \mw_{60}+\frac{392}{143 \sqrt{17}}  \mw_{80}\\

\hline 
T^-_1      & 1  &  1  &  \mw_{00} \\
               & 1  & 3  & -\frac{4 }{\sqrt{21}} \mw_{40} \\
                & 3  & 3   & \mw_{00}+\frac{6}{11} \mw_{40}+\frac{100}{33 \sqrt{13}} \mw_{60} \\
\hline 
T^+_1    & 4  & 4       & \mw_{00}+\frac{54}{143} \mw_{40}-\frac{4}{11 \sqrt{13}}  \mw_{60} -\frac{448}{143 \sqrt{17}}  \mw_{80}\\
    \hline

T^+_2 & 2  & 2 &  \mw_{00}-\frac{4}{7} \mw_{40} \\
    & 2  & 4 & -\frac{20\sqrt{3}}{77}  \mw_{40}+\frac{40}{11} \sqrt{\frac{3}{13}} \mw_{60} \\
    & 4 & 4 & \mw_{00}-\frac{54}{77} \mw_{40}+\frac{20 }{11 \sqrt{13}} \mw_{60} \\
\hline
T^-_2    & 3  & 3 & \mw_{00}-\frac{2}{11} \mw_{40}-\frac{60 }{11 \sqrt{13}} \mw_{60} \\
\bottomrule
\end{array}
$      
\end{tabular}
\caption{Non-zero reduced matrix elements in the cubic box
for integral angular momentum up to $J=4$  (symmetry group $O_{h}$).  
The matrix is symmetric in $J$ and $J'$ in each irrep-parity sector.}
\label{tab:me2Oh1}
\end{table}
\begin{table}
\begin{tabular}{c}
$      
\renewcommand{\arraystretch}{1.8}
\begin{array}{c |  c |  c | l }
\toprule
\Gamma &  J        & J'     & \mathcal{M}^\Gamma_{J l , J' l' }                  \\
\hline
G^\pm_1& {1\over 2}  &{1\over 2}    &  \mw_{00}                                                                               \\
   & {1\over 2}  & {7\over 2}  &  \mp \frac{4 }{\sqrt{21}} \mw_{40}  \\
   & {7\over 2}  & {7\over 2}  & \mw_{00}+\frac{6}{11} \mw_{40}+\frac{100 }{33 \sqrt{13}}\mw_{60} \\
\hline     
G^\pm_2 &  {5\over 2} & {5\over 2}     &  \mw_{00}-\frac{4}{7} \mw_{40}      \\

    &  {5\over 2}  & {7\over 2}    & \pm  \frac{20\sqrt{3}}{77}  \mw_{40} \mp \frac{40}{11} \sqrt{\frac{3}{13}} \mw_{60}  \\
    &  {7\over 2} & {7\over 2}    &  \mw_{00}-\frac{54}{77} \mw_{40}+\frac{20}{11 \sqrt{13}}  \mw_{60} \\

\hline
H ^\pm  &  {3\over 2}  & {3\over 2}    &   \mw_{00}       \\
    &  {3\over 2}  & {5\over 2}     &  \mp \frac{2 \sqrt{6}}{7} \mw_{40}        \\
    &  {3\over 2} & {7\over 2}    &   \frac{2}{7} \sqrt{\frac{10}{3}} \mw_{40} \\
    &  {5\over 2}& {5\over 2}     &   \mw_{00}+\frac{2}{7} \mw_{40}       \\
    &  {5\over 2} & {7\over 2}    & \mp \frac{12\sqrt{5}}{77}  \mw_{40} \mp \frac{20}{11} \sqrt{\frac{5}{13}} \mw_{60}  \\
    &  {7\over 2} & {7\over 2}    &  \mw_{00}+\frac{6}{77} \mw_{40}-\frac{80 }{33 \sqrt{13}} \mw_{60}   \\

\bottomrule
\end{array}
$      
\end{tabular}
\caption{Non-zero reduced matrix elements in the cubic box 
 for half-integral angular momentum up to $J=7/2$ (symmetry group $^2O_{h}$).  The even/odd parity sectors are indicated by the upper/lower signs.
The matrix is symmetric in $Jl$ and $J'l'$ in each irrep sector.}
\label{tab:me2Oh2}
\end{table}
%

\section{Relations between the cubic and elongated cases}

Having derived the phaseshift formulas in both the cubic and elongated boxes, we would like to explore the relationship between the two.
Going from cubic box to elongated box, the symmetry is reduced.
The two have different irreps:
$A^\pm_1, A^\pm_2, E^\pm, T^\pm_1, T^\pm_2, G^\pm_1, G^\pm_2, H^\pm$ 
with respective dimensionality $1,1,2,3,3,2,2,4$ in the cubic box,
and  $A^\pm_1, A^\pm_2, E^\pm, B^\pm_1, B^\pm_2, G^\pm_1, G^\pm_2$ 
with respective dimensionality $1,1,2,1,1,2,2$ in the elongated box.
How do they transition to one another?
The descent in symmetry follows the so-called subduction rules in group theory (see, for example, Ref.~\cite{Altmann:1994}), 
shown in Table~\ref{tab:sub}. 
In the same table, the rules for $^2C_{4v}$  are also given. They are relevant for states with non-zero momentum, to be discussed in the next section. 
\begin{table}
\begin{tabular}{c}
$      
\renewcommand{\arraystretch}{1.8}
\begin{array}{c | c | c | c | c | c | c | c | c | c | c | c | c | c | c | c | c }
\toprule  
^2O_h     & A^+_1 & A^+_2 & E^+ & T^+_1 & T^+_2 & G^+_1 & G^+_2 & H^+ & A^-_1 & A^-_2 & E^- & T^-_1 & T^-_2 & G^-_1 & G^-_2 & H^- \\
\hline   
^2D_{4h} & A^+_1 & B^+_1 & A^+_1\oplus B^+_1 & A^+_2\oplus E^+ & B^+_2\oplus E^+ & G^+_1 & G^+_2 & G^+_1\oplus G^+_2  
                & A^-_1 & B^-_1 & A^-_1\oplus B^-_1 & A^-_2\oplus E^- & B^-_2\oplus E^- & G^-_1 & G^-_2 & G^-_1 \oplus G^-_2 \\
\hline   
^2C_{4v} & A_1 & B_1 & A_1\oplus B_1 & A_2\oplus E & B_2\oplus E & G_1 & G_2 & G_1\oplus G_2 
                & A_2 & B_2 & A_2\oplus B_2 & A_1\oplus E & B_1\oplus E & G_1 & G_2 & G_1\oplus G_2\\
\bottomrule
\end{array}
$      
\end{tabular}
\caption{Subduction rules in the descent in symmetry in the group chain from the cubic box ($^2O_h$), 
to the elongated box ($^2D_{4h}$), to moving frame ($^2C_{4v}$).}
\label{tab:sub}
\end{table}

Specifically, we want to explore the relationships manifested in the matrix elements  
$\mathcal{M}^\Gamma_{J ln , J' l'n' } $ for phaseshifts.
In the case of integer angular momentum, 
we find the following correspondence by comparing the matrix elements in Table~\ref{tab:me2Oh1} and  Table~\ref{tab:me2D4h1}.
In the following, the cubic elements are on the left-hand side, the elongated  on the right-hand side.
\begin{enumerate}[(i)]
\item
For $(J,J')=(0,0)$, the $A_1$ has  one-to-one correspondence  
\beq \mm^{A^+_1}_{00} = \mm^{A^+_1}_{00}. \eeq
\item
For $(J,J')=(1,1)$, $T^-_1$ splits into $A^-_2$ and  $E^-$ according to 
\beq \mm^{T^-_1}_{11} =  {1\over 3}  \mm^{A^-_2}_{11}  +  {2\over 3} \mm^{E^-}_{11}. \label{eq:t1} \eeq
The factors are related to the fact that
$T_1$ is a three-dimensional irrep, whereas $A_2$ one-dimensional and $E$ two-dimensional.
Going from the elongated to the cubic symmetry, even though the matrix elements $\mm^{A^-_2}_{11}$ and $\mm^{E^-}_{11}$ 
will individually go to $\mm^{T^-_1}_{11}=\mw_{00}$ in the limit $\eta\rightarrow 1$ because $\mw_{20}$ goes to zero in the same limit, Eq.~\ref{eq:t1} shows how to follow this limit by subduction rule  in this particular channel. If coupling to $J=3$ and higher can be ignored, this relationship translates directly into one for the phaseshift:
\beq \cot\delta_1(\mathcal{T}^-_1) =  {1\over 3} \cot\delta_1(A^-_2) +  {2\over 3} \cot\delta_1(E^-). \eeq

\item
For $(J,J')=(2,2)$, there are two scenarios.  
The first is that $E^+$ splits into $A^+_1$ and  $B^+_1$ according to 
\beq  \mm^{E^+}_{22} = {1\over 2}(\mm^{A^+_1}_{22} + \mm^{B^+_1}_{22}). \eeq
The second is that $T^+_2$ splits into $B^+_2$ and $E^+$ according to 
\beq \mm^{T^-_1}_{22} =  {1\over 3}  \mm^{B^+_2}_{22}  +  {2\over 3} \mm^{E^+}_{22}. \eeq
In both cases, the condition $\mw_{44} = {\sqrt{70}\over 14}  \mw_{40}$ in the cubic box has been used. 
\end{enumerate}
Relations for other $JJ'$ combinations can be found in similar fashion.

In the case of half-integral angular momentum, 
we find the following correspondence by comparing the matrix elements in Table~\ref{tab:me2Oh2} and  Table~\ref{tab:me2D4h2}.
\begin{enumerate}[(i)]
\item
For $(J,J')=(1/2,1/2)$, there is one-to-one correspondence  
\beq \mm^{G^\pm_1}_{{1\over 2}{1\over 2}} = \mm^{G^\pm_1}_{{1\over 2}{1\over 2}}. \eeq
\item
For $(J,J')=(3/2,3/2)$, $H^\pm$ splits into $G^\pm_1$ and $G^\pm_2$ evenly according to 
\beq \mm^{H^\pm}_{{3\over 2}{3\over 2}} = {1\over 2} \left(\mm^{G^\pm_1}_{{3\over 2}{3\over 2}} + \mm^{G^\pm_2}_{{3\over 2}{3\over 2}}\right). \eeq
The dimensionality also checks out ($4=2\oplus 2$).
\item
For $(J,J')=(5/2,5/2)$, there are two scenarios.  
The first is $G^\pm_2$ to $G^\pm_2$, but the latter has a two-fold multiplicity labeled by $n,n'$ in the notation $\mm^{\Gamma}_{JJ'}(n,n')$.  The specific combination is found to be 
\beq
\mm^{G^\pm_2}_{{5\over 2}{5\over 2}}  =  {5\over 6} \mm^{G^\pm_2}_{{5\over 2}{5\over 2}} (1,1) - {\sqrt{5}\over 6}[ \mm^{G^\pm_2}_{{5\over 2}{5\over 2}}(1,2)+ \mm^{G^\pm_2}_{{5\over 2}{5\over 2}}(2,1)] + {1\over 6} \mm^{G^\pm_2}_{{5\over 2}{5\over 2}}(2,2). 
\eeq
The second is that $H^\pm$ splits into $G^\pm_1$ and $G^\pm_2$ according to 
\beq
\mm^{H^\pm}_{{5\over 2}{5\over 2}}  =  {1\over 2} \mm^{G^\pm_1}_{{5\over 2}{5\over 2}}
+ {1\over 12} \mm^{G^\pm_2}_{{5\over 2}{5\over 2}} (1,1) + {\sqrt{5}\over 12} [\mm^{G^\pm_2}_{{5\over 2}{5\over 2}}(1,2)+ \mm^{G^\pm_2}_{{5\over 2}{5\over 2}}(2,1)] + {5\over 12} \mm^{G^\pm_2}_{{5\over 2}{5\over 2}}(2,2). 
\eeq


\end{enumerate}
Relations for other $JJ'$ combinations can be found in similar fashion.

\section{Moving states in a cubic box}

So far we have considered two-body states that are at rest; 
the two particles have back-to-back nonzero momentum, but the total momentum $\bm P=0$ in the lab frame.
Now we consider giving the system a boost. 
In the center-of-mass frame (CM) the cubic box becomes a parallelepiped,
in which the side parallel to the directions of the boost is contracted by the Lorentz boost factor $\gamma$,
whereas the size in the perpendicular direction is unchanged.
The advantage of boosting is that it can lower the center-of-mass energy, thus allowing wider access to the resonance region.
The invariant energy of the system is
\beq W=\sqrt{m^2_1+\bm k^2}+\sqrt{m^2_2+\bm k^2}, \label{eq:winv}\eeq
 where $\bm k$ is the relative momentum in the CM frame.
The energy in the lab frame 
\beq E_{lab}=\sqrt{m^2_1+\bm p^2_1}+\sqrt{m^2_2+\bm p^2_2},  \label{eq:elab} \eeq
is the same as $W$ when the system is at rest.
But when the system is moving with momentum 
\beq \bm P=\bm p_1 + \bm p_2={2\pi \over L} \bm d \eeq
as measured in the lab frame,
the CM energy is lowered according to
\beq
W=\sqrt{E_{lab}^2-\bm P^2}.
\eeq
The two energies are related by $W=E_{lab}/\gamma$ where the boost factor $\gamma$ and boost velocity 
$\bm v$ are given by
\beq \gamma={1\over \sqrt{1-\bm v^2}} \text{  with  } \bm v = {\bm P \over E_{lab}}.  \label{eq:gamv} \eeq
The procedure to extract phaseshift is to first measure the interaction energy $E_{lab}$, then determine $W$ via the boost factor, 
then $\bm k$ via Eq.~\ref{eq:winv}, then $\delta(k)$ via the L\"{u}scher formula.
Ref.~\cite{Gockeler:2012yj} considered boosts in three different directions and Ref.~\cite{Leskovec:2012gb} in
two different directions. 
We restrict ourselves to boosts only in the $z$-direction: $\bm d=(0,0,d_z)$ with $d_z\in \mathds Z$.
These boosts preserve the symmetry of the elongated box, as viewed from the CM frame.
The wavefunction for the relative position $\psi(\bm r)$ still satisfies the Helmholtz equation,
Eq.~\ref{eq:hemholtz},
when the distance between particles is larger than the interaction range, but the
boundary conditions are different for boosted states~\cite{Rummukainen:1995vs,Fu:2011xz, Leskovec:2012gb,Gockeler:2012yj}:
\beq
\psi(\bm r+\hat\gamma\bm n L) = e^{i\pi A \bm n \cdot \bm d} \psi(\bm r) \,,
\eeq
where $\hat\gamma\bm n\equiv \gamma \bm n_\parallel + \bm n_\perp$ and
\begin{equation}
A=1+ {m_1^2-m_2^2 \over W^2}\,.
\end{equation}
The full symmetry group that preserves these boundary conditions for the case when $\bm d$ is
parallel with the $z$-direction in the cubic box is $^2C_{4v}$. 
We relegate the details for this group to Appendix~\ref{app:boost}.
The most important difference from the zero momentum case is that the boundary conditions specified
above are not invariant under parity, so the solutions of the Helmholtz equation are a mixture 
of different parities. For the meson-baryon states, this means that we can no longer identify
the orbital angular momentum $l$ for a given $J$ using parity. For the two-meson states, the
most important consequence is that the irreps now overlap with all angular momenta $J$, not just
the even (or odd) ones.

When the phaseshift  reduction is carried out for boosting using the basis vectors in Table~\ref{tab:basis2C4v},
the matrix elements are obtained in Table~\ref{tab:me2C4v1} for integer $J$ and Table~\ref{tab:me2C4v2} for half-integer $J$. 
For diagonal $JJ'$ and $ll'$ combinations, the two elements corresponding to the even and odd $l$ are  the same, so only one of them is shown. 
For example, $\mathcal{M}^\Gamma_{{1\over 2}0, {1\over 2}0}=\mathcal{M}^\Gamma_{{1\over 2}1, {1\over 2}1}$, 
$\mathcal{M}^\Gamma_{{3\over 2}1, {3\over 2}1}=\mathcal{M}^\Gamma_{{3\over 2}2, {3\over 2}2}$, and so on.
Notice the appearance of $\mw$ functions with odd values of $J$ due to loss of parity.
Our results agree fully with those in Ref.~\cite{Gockeler:2012yj} in the case of $\bm d=(0,0,1)$ for up to $J=2$ 
for mesons and up to $J=3/2$ for baryons, after accounting for the $\sqrt{2l+1}$ factor in the definition of the 
$\mw_{lm}$ functions. Our results extend to $J=3$ and $J=5/2$. Our results for integer $J$ in Table~\ref{tab:me2C4v1}  also largely agree with those in Ref.~\cite{Fu:2011xz}, except for the missing $\sqrt{3}$ factor in $\mw_{10}$ that is 
also pointed out in Ref.~\cite{Leskovec:2012gb}. The agreements give a concrete demonstration 
that the phaseshift formulas are independent of the basis vectors since we used different basis vectors. 

The $\mw$ functions in Table~\ref{tab:me2C4v1} and Table~\ref{tab:me2C4v2} are the modified versions 
of the cubic ones to incorporate boosting (indicated by $\gamma$ and $\bm d$) in the cubic box (no $\eta$ factor),
\beq
\mw_{lm}(1,q^2,\gamma)= \frac{\mathcal{Z}^{\bm d}_{lm}(1,q^2,\gamma)}{\pi^{\frac{3}{2}}\gamma q^{l+1}}\,.
\label{eq:wfun_boost}
\eeq
The relevant zeta function is
\beq
\mathcal{Z}_{lm}^{\bm d}(s,q^2,\gamma) = \sum_{ \widetilde{\bm n}\in P_{\bm d}(\gamma)} \frac{\mathcal{Y}_{lm}(\widetilde{\bm n})}{(\widetilde{\bm n}^2-q^2)^s},
\label{eq:zfun_boost}
\eeq
where the summation region changes to
\beq
P_{\bm d}(\gamma) =\left\{\widetilde{\bm n}\in\reals^3 \mid \widetilde{\bm n}=\hat{\gamma}^{-1}(\bm m+\frac{1}{2}A\, \bm d), \bm m\in \mathds{Z}^3 \right\},
\label{eq:psum_boost}
\eeq
where $\widetilde{\bm n}$ is over real numbers and $\bm m$ over integers.
The projector $\hat{\gamma}^{-1}$ operating on a vector $\bm n$ is defined as only affecting the boost direction
\beq
\hat{\gamma}^{-1} \bm n = {\bm n_{\parallel}\over \gamma} +\bm n_{\perp}
\text{  where  } \bm n_{\parallel} = {\bm v(\bm n\cdot\bm v) \over \bm v^2}   
\text{  and  }\ \bm n_{\perp}=\bm n - \bm n_{\parallel}.
\label{eq:gamma}
\eeq

The fact that the two particles have unequal masses enters explicitly in the zeta function through the scaling factor in front of the boost $\bm d$ in Eq.~\ref{eq:psum_boost}~\cite{Gockeler:2012yj,Leskovec:2012gb}, A, which reduces to one when the two masses are equal. 
For unequal masses, what happens if they are switched? 
This question can be answered by examining the mass dependence in the zeta function 
in Eq.~\ref{eq:zfun_boost}. 
Interchanging $m_1$ and $m_2$ only affects the $A$ factor in the summation grid in  Eq.~\ref{eq:psum_boost}. 
The result is a change in sign of the set of points to be summed over from $\widetilde{\bm n}$ to $- \widetilde{\bm n}$ (the mirror image grid).  
This leads to an overall sign change in the zeta function, which does not affect the phaseshift determinants. 
So the order of $m_1$ and $m_2$ does not matter; they have the same phaseshift formula.
There is a real physical effect, however,  that depends on the order of $m_1$ and $m_2$.
Let us use an example and assume $m_2 > m_1$.
For a given boost $\bm d=(0,0,1)$, 
there are two possible arrangements for momenta in the lab frame:
case 1 is $\bm p_1=(0,0,0)$ and  $\bm p_2=(0,0,2\pi/L)$, case 2 is $\bm p_1=(0,0,2\pi/L)$ and $\bm p_2=(0,0,0)$. According to Eq.~\ref{eq:elab}, when the heavier particle ($m_2$) possesses the nonzero momentum (case 1), the energy $E_{lab}$ is  lower than in case 2. 
It means case 1 has a lower CM energy $W$, which means a lower CM back-to-back momentum $\bm k$.
Another way of looking at it is by Eq.~\ref{eq:gamv}: case 1 has higher boost velocity $\bm v$ and boost factor $\gamma$, thus lower energy.

In the case of equal masses, parity (or space inversion) is restored, leading to considerable 
simplifications in Table~\ref{tab:me2C4v1} and Table~\ref{tab:me2C4v2}. 
All the odd $l$ zeta functions vanish. The matrix elements can then be separated into sectors by irrep-parity combinations.

In the absence of a boost ($\bm d=0$ and $\gamma=1$), the zeta function returns to the original one in the 
cubic box. How the cubic symmetry is restored follows the subduction rules between $^2O_h$ and $^2C_{4v}$  in Table~\ref{tab:sub},
similar to between $^2O_h$ and $^2D_{4h}$ discussed in the previous section, 
except the role reversals in $A_1$ and $B_1$ and $A_2$ and $B_2$ for negativity parity.

\begin{table*}[p]
\begin{tabular}{c}
$      
\renewcommand{\arraystretch}{2.0}
\begin{array}{c| c c | c c | c }
\toprule
\Gamma & J  & n  &  J'    & n'  & \mathcal{M}^\Gamma_{J n , J' n'}   \\
\hline 
A_1  & 0 & 1  & 0 & 1  & \mw_{00}  \\
    & 0 & 1  & 1 & 1  & i \mw_{10} \\
    & 0 & 1  & 2 & 1  & -\mw_{20} \\
    & 0 & 1  & 3 & 1  & -i \mw_{30} \\
    & 1 & 1  & 1 & 1  & \mw_{00}+\frac{2 }{\sqrt{5}} \mw_{20}\\
    & 1 & 1  & 2 & 1  & \frac{2 i }{\sqrt{5}} \mw_{10}+3 i \sqrt{\frac{3}{35}} \mw_{30} \\
    & 1 & 1  & 3 & 1  & -3 \sqrt{\frac{3}{35}} \mw_{20}-\frac{4 }{\sqrt{21}}  \mw_{40} \\
    & 2 & 1  & 2 & 1  & \mw_{00}+\frac{2\sqrt{5}}{7}  \mw_{20}+\frac{6}{7} \mw_{40}  \\
    & 2 & 1  & 3 & 1  & 3 i \sqrt{\frac{3}{35}} \mw_{10}+\frac{4 i }{3 \sqrt{5}} \mw_{30}+\frac{10}{3} i \sqrt{\frac{5}{77}} \mw_{50}  \\
    & 3 & 1  & 3 & 1  & \mw_{00}+\frac{4} {3 \sqrt{5}} \mw_{20}+\frac{6}{11} \mw_{40}+\frac{100 }{33 \sqrt{13}} \mw_{60} \\
\hline 
B_{1/2}  & 2 & 1  & 2 & 1  & \mw_{00}-\frac{2\sqrt{5}}{7}  \mw_{20}+\frac{1}{7} \mw_{40} \pm \sqrt{\frac{10}{7}} \mw_{44} \\
                & 2 & 1  & 3 & 1  & i \sqrt{\frac{3}{7}} \mw_{10}-\frac{2i}{3}  \mw_{30}+\frac{5 i }{3 \sqrt{77}}\mw_{50} \pm i \sqrt{\frac{10}{11}} \mw_{54}  \\
                & 3 & 1  & 3 & 1  &\mw_{00}-\frac{7}{11} \mw_{40}+\frac{10 }{11 \sqrt{13}}\mw_{60} \pm\frac{\sqrt{70}}{11}  \mw_{44}\pm\frac{10}{11} \sqrt{\frac{14}{13}} \mw_{64} \\
\hline

E & 1 & 1 & 1 & 1 & \mw_{00} - {1\over \sqrt{5}} \mw_{20}           \\
    & 1 & 1 & 2 & 1 & i \sqrt{\frac{3}{5}} \mw_{10}-\frac{3 i }{\sqrt{35}} \mw_{30}   \\
    & 1 & 1 & 3 & 1 & -3 \sqrt{\frac{2}{35}} \mw_{20}+\sqrt{\frac{2}{7}} \mw_{40}   \\
    & 1 & 1 & 3 & 2 &  \frac{2 }{\sqrt{3}}  \mw_{44} \\
 
    & 2 & 1 & 2 & 1 & \mw_{00}+\frac{\sqrt{5} }{7} \mw_{20}-\frac{4}{7} \mw_{40}      \\
    & 2 & 1 & 3 & 1 & 2 i \sqrt{\frac{6}{35}} \mw_{10}+\frac{1}{3} i \sqrt{\frac{2}{5}} \mw_{30}-\frac{5}{3} i \sqrt{\frac{10}{77}} \mw_{50}   \\
    & 2 & 1 & 3 & 2 & -2 i \sqrt{\frac{5}{33}}  \mw_{54}  \\

    & 3 & 1 & 3 & 1 & \mw_{00}+\frac{1}{\sqrt{5}}\mw_{20}+\frac{1}{11} \mw_{40}-\frac{25 }{11 \sqrt{13}} \mw_{60} \\
    & 3 & 2 & 3 & 2 & \mw_{00}-\frac{\sqrt{5} }{3} \mw_{20}+\frac{3}{11} \mw_{40}-\frac{5}{33 \sqrt{13}}  \mw_{60} \\
\bottomrule
\end{array}
$      
\end{tabular}
\caption{Non-zero reduced matrix elements for boosting in the cubic box ($C_{4v}$ symmetry group)
for integral angular momentum up to $J=3$.  
The $A_2$ sector does not appear below $J=4$.
The $B_{1}$ and $B_2$ irreps are combined as indicated by the upper/lower signs.
The matrix is hermitian in $Jn$ and $J'n'$ in each irrep sector.
}
\label{tab:me2C4v1}
\end{table*}
\begin{table}
\begin{tabular}{c}
$      
\renewcommand{\arraystretch}{1.8}
\begin{array}{c | c c c | c c c | c }
\toprule
\Gamma & J              & l &&  J'         & l' &&  \mathcal{M}^\Gamma_{J l , J' l' }                                                          \\
\hline                                             
G_1    & {\bm 1\over 2} & 0 &&  {1\over 2} & 0  &&  \mw_{00}                                                                                        \\
       & {1\over 2}     & 0 &&  {1\over 2} & 1  &&  -\frac{i }{\sqrt{3}}    \mw_{10}                                                             \\
\hline                                             
       & {1\over 2}     & 0 &&  {3\over 2} & 1  &&  i \sqrt{\frac{2}{3}} \mw_{10}                                                                     \\
       & {1\over 2}     & 0 &&  {3\over 2} & 2  &&  \sqrt{\frac{2}{5}} \mw_{20}                                                                       \\
       & {1\over 2}     & 1 &&  {3\over 2} & 1  &&  -\sqrt{\frac{2}{5}} \mw_{20}                                                                      \\
       & {1\over 2}     & 1 &&  {3\over 2} & 2  &&  i \sqrt{\frac{2}{3}} \mw_{10}                                                                     \\
\hline                                             
       & {3\over 2}     & 1 &&  {3\over 2} & 1  &&  \mw_{00}+\frac{1}{\sqrt{5}}  \mw_{20}                                                                \\
       & {3\over 2}     & 1 &&  {3\over 2} & 2  &&  - i \frac{1 }{5 \sqrt{3}} \mw_{10} - i \frac{9 }{5 \sqrt{7}} \mw_{30}                                     \\
\hline                                             
       & {1\over 2}     & 0 &&  {5\over 2} & 2  &&  -\sqrt{\frac{3}{5}} \mw_{20}                                                                      \\
       & {1\over 2}     & 0 &&  {5\over 2} & 3  &&  i \sqrt{\frac{3}{7}} \mw_{30}                                                                     \\
       & {1\over 2}     & 1 &&  {5\over 2} & 3  &&  -i \sqrt{\frac{3}{7}} \mw_{30}                                                                    \\
\hline                                             
       & {3\over 2}     & 1 &&  {5\over 2} & 2  && i \frac{3\sqrt{2}}{5}   \mw_{10}+\frac{2}{5} i \sqrt{\frac{6}{7}} \mw_{30}                           \\
       & {3\over 2}     & 1 &&  {5\over 2} & 3  &&  \frac{1}{7} \sqrt{\frac{6}{5}} \mw_{20}+\frac{2\sqrt{6}}{7}  \mw_{40}                               \\
       & {3\over 2}     & 2 &&  {5\over 2} & 2  &&  -\frac{1}{7} \sqrt{\frac{6}{5}} \mw_{20}-\frac{2\sqrt{6}}{7}  \mw_{40}                              \\
\hline                                             
       & {5\over 2}     & 2 &&  {5\over 2} & 2  &&  \mw_{00}+\frac{8 }{7 \sqrt{5}}\mw_{20} +\frac{2}{7} \mw_{40}                                           \\
       & {5\over 2}     & 2 &&  {5\over 2} & 3  &&  -i\frac{\sqrt{3}}{35}   \mw_{10} - i \frac{8 }{15 \sqrt{7}}\mw_{30} - i \frac{50 }{21 \sqrt{11}}  \mw_{50}\\

\hline  
    & J          & l & n & J'         & l' & n'& \mathcal{M}^\Gamma_{J l n , J' l' n' }                                                          \\
\hline                                       
G_2 & {3\over 2} & 1 & 1 & {3\over 2} & 1 & 1   & \mw_{00}-\frac{1}{\sqrt{5}}  \mw_{20}                                                                \\
    & {3\over 2} & 1 & 1 & {3\over 2} & 2 & 1   & - i \frac{\sqrt{3}}{5}  \mw_{10}+ i \frac{3}{5 \sqrt{7}}     \mw_{30}                                \\
\hline                                      
    & {3\over 2} & 1 & 1 & {5\over 2} & 2 & 1   & i \frac{2\sqrt{3} }{5} \mw_{10}- i \frac{6 }{5 \sqrt{7}}   \mw_{30}                                  \\
    & {3\over 2} & 1 & 1 & {5\over 2} & 3 & 1   & \frac{6 }{7 \sqrt{5}} \mw_{20} - \frac{2}{7} \mw_{40}                                                  \\
    & {3\over 2} & 1 & 1 & {5\over 2} & 3 & 2   & -2 \sqrt{\frac{2}{7}} \mw_{44}                                                                    \\
    & {3\over 2} & 2 & 1 & {5\over 2} & 2 & 1   & -\frac{6 }{7 \sqrt{5}} \mw_{20} +\frac{2}{7} \mw_{40}                                                 \\
    & {3\over 2} & 2 & 1 & {5\over 2} & 2 & 2   & 2 \sqrt{\frac{2}{7}} \mw_{44}                                                                     \\
\hline                                      
    & {5\over 2} & 2 & 1 & {5\over 2} & 2 & 1   & \mw_{00}+\frac{2 }{7 \sqrt{5}} \mw_{20} - \frac{3}{7} \mw_{40}                                           \\
    & {5\over 2} & 2 & 1 & {5\over 2} & 2 & 2   & \sqrt{\frac{2}{7}} \mw_{44}                                                                       \\
    & {5\over 2} & 2 & 2 & {5\over 2} & 2 & 2   & \mw_{00}-\frac{2 \sqrt{5} }{7}\mw_{20}+\frac{1}{7} \mw_{40}                                           \\
    & {5\over 2} & 2 & 1 & {5\over 2} & 3 & 1   & - i \frac{3 \sqrt{3} }{35} \mw_{10} - i \frac{2 \sqrt{7} }{15}\mw_{30}+ i \frac{25 }{21 \sqrt{11}} \mw_{50} \\
    & {5\over 2} & 2 & 1 & {5\over 2} & 3 & 2   & 5 i \sqrt{\frac{2}{77}} \mw_{54}                                                        \\
    & {5\over 2} & 2 & 2 & {5\over 2} & 3 & 2   & i \frac{\sqrt{3}}{7}  \mw_{10}- i \frac{2 }{3 \sqrt{7}} \mw_{30}+ i \frac{5 }{21 \sqrt{11}} \mw_{50}    \\

\bottomrule
\end{array}
$          
\end{tabular}
\caption{Non-zero reduced matrix elements for boosting in the cubic box ($^2C_{4v}$ symmetry group)
 for half-integral angular momentum up to $J=5/2$.  There is no multiplicity in the $G_1$ sector, but 
 two-fold multiplicity in the $G_2$ sector. The horizontal lines separate different combinations of $JJ'$.
The matrix is hermitian in $Jln$ and $J'l'n'$ in each irrep sector.}
\label{tab:me2C4v2}
\end{table}
%

\section{Moving states in an elongated box}

Elongation in one of the dimensions picks out a special direction in space; so does boosting. 
The general situation when the two directions do not align is complicated.
However, when the elongation and the boost are in the same direction, such as the $z$-axis considered in this work, deriving the L\"uscher formulas for moving states is considerably simpler.

The key observation is that the $^2C_{4v}$ symmetry group for moving states is isomorphic to $^2D_{4}$ of the $z$-elongated box. 
Therefore, the matrix elements for the phaseshifts have exactly the same forms as those in Table~\ref{tab:me2C4v1} and Table~\ref{tab:me2C4v2}. 
The only difference is that we need to make a small change in the zeta functions,
namely, add the elongation factor $\eta$  in Equations~\ref{eq:wfun_boost} to~\ref{eq:psum_boost}, so they now read
\beq
\mw_{lm}(1,q^2,\gamma,\eta)= \frac{\mathcal{Z}^{\bm d}_{lm}(1,q^2,\gamma,\eta)}{\pi^{\frac{3}{2}}\eta \gamma q^{l+1}},
\label{eq:wfun_boost_eta}
\eeq
and
\beq
\mathcal{Z}_{lm}^{\bm d}(s,q^2,\gamma,\eta) = \sum_{ \widetilde{\bm n}\in P_{\bm d}(\gamma,\eta)} \frac{\mathcal{Y}_{lm}(\widetilde{\bm n})}{(\widetilde{\bm n}^2-q^2)^s},
\label{eq:zfun_boost_eta}
\eeq
where the summation grid changes to
\beq
P_{\bm d}(\gamma,\eta) =\left\{\widetilde{\bm n}\in\reals^3 \mid \widetilde{\bm n}=\hat{\gamma}^{-1}\hat{\eta}^{-1}(\bm m+\frac{1}{2}A\, \bm d), \bm m\in \mathds{Z}^3 \right\},
\label{eq:psum_boost_eta}
\eeq
with the projector $\hat{\eta}^{-1}$ acting on a vector $\bm m$
 to mean $\hat{\eta}^{-1} \bm m =(m_x,m_y,m_z/\eta)$.
Since the boost and elongation are in the same $z$-direction, 
the factors always appear as a product $\gamma\eta$ 
in the zeta function, facilitating its evaluation.

Due to lack of parity in boosted states, there is mixing between odd and even $J$ and the entire sector for each irrep becomes coupled.
This means that the phaseshift formulas are generally more complicated for moving states than for the ones at rest.
For example, the $A_2$ irrep is no longer a good sector for isolating $J=1$ resonances since $J=0$ appears below it 
as the ground state. The $E$ sector still has $J=1$ as the ground state, but mixes with all states $J=2$ and higher, 
as opposed to only odd states $J=3, 5, \ldots$, as is the case for states at rest.
If we only consider coupling with $J=2$, the phaseshift formula is
\beq
\left |
\begin{array}{cc}       
 \mw_{00} - {1\over \sqrt{5}} \mw_{20}   - \cot\delta_{1} &   i \sqrt{\frac{3}{5}} \mw_{10}-\frac{3 i }{\sqrt{35}} \mw_{30}      \\
-i \sqrt{\frac{3}{5}} \mw_{10} + \frac{3 i }{\sqrt{35}} \mw_{30}          & \mw_{00}+\frac{\sqrt{5} }{7} \mw_{20}-\frac{4}{7} \mw_{40}   - \cot\delta_{2}   \\
\end{array}             
\right |                
=0.                     
\label{eq:detE1boost}
\eeq
This formula applies to processes such as $\pi K\rightarrow K^*$, and 
$\pi \rho\rightarrow a_1$ (S-wave only). 

In the $G_1$ sector, the channels $\delta_{{1\over 2}0}$ ($S_{11}$ resonance for $\pi$-$N$ scattering) 
and $\delta_{{1\over 2}1}$ ($P_{11}$ or Roper  resonance) become coupled,
\beq
\left |
\begin{array}{cc}       
\mw_{00} - \cot\delta_{{1\over 2}0} &  -\frac{i }{\sqrt{3}}    \mw_{10}     \\
 \frac{i }{\sqrt{3}}    \mw_{10}                & \mw_{00} - \cot\delta_{{1\over 2}1}   \\
\end{array}             
\right |                
=0,                     
\label{eq:detG1boost}
\eeq
if we ignore mixing with higher $J$.
So the determination of spin-1/2 resonances requires $\mw_{10}$ in addition to $\mw_{00}$ at the leading order.
The two phaseshifts have to be extracted simultaneously.

In the $G_2$ sector,  $\delta_{{3\over 2}1}$ and $\delta_{{3\over 2}2}$ become similarly coupled,
\beq
\left |
\begin{array}{cc}       
\mw_{00}-\frac{1}{\sqrt{5}}  \mw_{20}   - \cot\delta_{{3\over 2}1} &  - i \frac{\sqrt{3}}{5}  \mw_{10}+ i \frac{3}{5 \sqrt{7}}     \mw_{30}      \\
i \frac{\sqrt{3}}{5}  \mw_{10}- i \frac{3}{5 \sqrt{7}}     \mw_{30}        &\mw_{00}-\frac{1}{\sqrt{5}}  \mw_{20}   - \cot\delta_{{3\over 2}2}   \\
\end{array}             
\right |                
=0,                     
\label{eq:detG2boost}
\eeq
if we ignore mixing with higher $J$. So the determination of $\delta_{{3\over 2}1}$ ($\Delta$ or $P_{33}$ resonance ) requires  four  zeta functions ($\mw_{l0}$ with $l=0,1,2,3$) and the knowledge of $\delta_{{3\over 2}2}$
($D_{33}$ resonance). Only when the $\delta_{{3\over 2}2}$ contribution can be ignored do we get the simple formula
\beq
 \cot\delta_{{3\over 2}1}  = \mw_{00}-\frac{1}{\sqrt{5}}  \mw_{20}. 
\eeq

The above discussion applies to the general case of unequal masses.
In the case of equal masses, parity is restored, and the tables simplify considerably.
All the odd $l$ zeta functions vanish and the matrix elements can be separated by irrep and parity.
The same result can be reached by adding the boost directly to the results derived in the elongated box 
under the symmetry group $^2D_{4h}$, given in Table~\ref{tab:me2D4h1} and Table~\ref{tab:me2D4h2}.
This perspective was adopted in Ref.~\cite{Guo:2016zos}, but it is only valid for equal masses.

Finally, we point out that in the limit when the boost goes to zero we recover the 
symmetry of the states at rest in elongated boxes. To see this just follow the subduction 
rules between $^2D_{4h}$ and $^2C_{4v}$  in Table~\ref{tab:sub}.
Indeed, Table~\ref{tab:me2C4v1} and Table~\ref{tab:me2C4v2} go back to Table~\ref{tab:me2D4h1} and Table~\ref{tab:me2D4h2} 
after turning off the odd-$l$ zeta functions, setting $\bm d=0$ and $\gamma=1$ in the remaining ones,  and re-organizing by irrep and parity.

\section{Conclusion}

We have derived  L\"{u}scher phaseshift formulas for two-body elastic scattering in
elongated boxes. We analyzed two scenarios: scattering of spinless mesons and
scattering of a spin zero meson from a spin-$1/2$ baryon. For each of these
scenarios we discussed the case where the two-body state is at rest with respect
to the box, or moving along the elongated direction.

Our interest in elongated boxes stems
from the fact that they allow us to vary the geometry of the box, and consequently
the kinematics, with minimal amount of computer resources. On the other hand,
elongated boxes have a different symmetry group than the cubic case and this
has to be taken into account when designing interpolators and when connecting
the infinite volume phase-shifts with the two-body energies. 

The main goal of this study was to derive the relevant L\"uscher formulas for
$\pi$-$N$ scattering on elongated boxes. 
To derive these relations we followed the methods developed for cubic boxes,
while accounting for the different symmetry of our setup. 
The formulas derived for baryon-meson scattering in elongated boxes are, to our
knowledge, new both for the moving and at-rest states. For the meson-meson
scattering the formulas derived here for the states at rest agree with the one derived by 
in Ref.~\cite{Feng:2004ua}, while for the moving case they are mostly 
new. The only moving case considered in the literature was the $A_2^-$ case for $\pi$-$\pi$
scattering~\cite{Guo:2016zos} and our results agree.


%
%
%

Elongated boxes offer a cost-effect way of varying the box size. 
The sensitivity of the energy spectrum to the elongation factor $\eta$  is a channel-dependent problem. For an example in 
the $\pi\pi\rightarrow \rho$ channel, see our previous work in Ref.~\cite{Guo:2016zos}. Generically, we expect these L\"{u}scher phaseshift formulas to be valid up to corrections on the order of $e^{-m_\pi L}$. 

As a validation of these formulas, we re-derived the results in the cubic box using the same approach that treats single and double-cover groups in a unified manner, as detailed in the appendices. How the symmetry is restored from the elongated box to the cubic box is governed by subduction rules and examples are given.

Boosting of the two-particle system in both the cubic and elongated boxes allows lower energies to be accessed, thus a wider coverage of the resonance region. 
The trade-off is the loss of parity for unequal masses which means mixing of even and odd states and generally more complicated formulas. We clarified the relationships between the various scenarios (cubic, elongated, and boosting) 
and how they transition from one another at the phaseshift level.

Finally, we note that the methods used here can be readily extended to other interesting cases.
One example is the $\pi$-$\rho$ scattering in the $a_1$ channel. Since $\rho$ is not
spinless, the formalism used here needs to be extended to a multi-channel one to account for the fact that the
S-wave and D-wave states mix. However, if we consider that the S-wave
channel dominates, some of the formulas derived here can be directly applied.
Efforts are under way to investigate the $a_1$ and 
the $\Delta $ resonances using elongated boxes in lattice QCD simulations.

\acknowledgements

This work is supported in part by the U.S. Department of Energy grant DE-FG02-95ER40907 and the National Science Foundation CAREER grant PHY-1151648. 
A.A. gratefully acknowledges 
the hospitality of the Physics Departments at the Universities of Maryland and Kentucky, 
and the Albert Einstein Center at the University of Bern where part
of this work was carried out.

\appendix

\newpage
\section{Symmetry group properties in the elongated box}
\label{app:elongated}

In this appendix and the ones that follow, we give an overview of the basic ideas and terminology of group theory 
adapted for the understanding of angular momentum and phaseshift reduction in the elongated box.
It is reasonably detailed for a coherent and self-contained picture.
The literature on group theory is vast, coming from many perspectives including solid-state physics, quantum chemistry, and mathematics.
We limit ourselves to a selected few~\cite{Tinkham:1992,Johnson:1982,Moore:2005dw,Altmann:1986,Altmann:1994,Chen:2002} that 
should be familiar to a student of physics.

In the elongated box (also known as a cuboid or square prism), 
the basic symmetry group involving only spatial rotations 
is called the dihedral group $D_4$. As far as group operations are concerned, 
the $D_4$ group is isomorphic to the symmetry of a square, with 8 simple elements (operations that 
return the square to itself). They can be divided into 5 conjugacy classes (operations that are equivalent): 
the identity (I), 
two $\pi/2$ rotations about the perpendicular z-axis ($2 C_4$), 
one $\pi$ rotation about the z-axis ($C_2$),  
two $\pi$  rotations about $x$ and $y$ axes ($2C'_2$), 
and  two $\pi$  rotations about the two diagonals  ($2C{''}_2$). 
The operations can be visualized in Fig.~\ref{fig:box_elongated}.
To fully describe the physics at hand, we need two extensions.
The box has a symmetry under space inversion (or parity) about the $xy$-plane, which entails the symmetry group $D_{4h}$. To describe half-integral angular momentum, the double-covered group of $D_4$, denoted as $^2D_4$, 
is required. So the full symmetry group in the elongated box will be called $^2D_{4h}$. 
All these variants will be explained below.

Given a group, a matrix representation can be constructed. The representations are generally not unique 
since a similarity transformation can lead to a different representation. What is unique is the {\em character} of each group element, defined as the trace of the matrix that represents the element. 
The character is the same within each conjugacy class.
Usually representations of a finite group are reducible, 
so we seek the set of  {\bf irreducible representations} (hereafter referred to as {\bf irreps}) for the group. One can think of the procedure as reducing a matrix in its block diagonal form by similarity transformations. For this reason, any representation of a finite group can be broken up into a direct sum of its irreps.
Finding the irreps of a group is one of the most important tasks in group theory, followed by how things transform under the irreps.

The $D_4$ group has 5 irreps conventionally named $A_1$,  $A_2$,  $B_1$, $B_2$, and $E$, 
with respective dimensions 1, 1, 1, 1, 2 (whose squares sum to 8). 
It is an example of a  general group property that the square of irrep dimensions sum 
to the total number of elements (or group order).  
Another useful property is that the number of irreps is equal to the number of conjugacy classes.

\begin{table*}                   
\begin{tabular}{c}               
$                                
\renewcommand{\arraystretch}{1.8}
\begin{array}{c | c | c | c | c | r | r | r | r | c | c | c}
\toprule                         
\text{Class\textbackslash Irrep} & k  & \bm n      & \omega           & (\alpha,\beta,\gamma)                  & A_1 & A_2 & B_1 & B_2 & E            & G_1                                        & G_2                                                  \\
\hline                           
\text{I}                         & 1  & \{0,0,1\}  & 4 \pi            & \{0,0,0\}                              & 1   & 1   & 1   & 1   & 1            & 1                                          & 1                                                    \\
\hline                           
\widetilde{I}                    & 2  & \{0,0,1\}  & 2 \pi            & \{0,0,2 \pi \}                         & 1   & 1   & 1   & 1   & 1            & -1                                         & -1                                                   \\
\hline                           
C_{4z}^+                         & 3  & \{0,0,1\}  & \frac{\pi }{2}   & \left\{0,0,\frac{\pi }{2}\right\}      & 1   & 1   & -1  & -1  & -i \sigma _2 & \frac{1-i                                  
                                                                                                                                                       \sigma _2}{\sqrt{2}}                        & \frac{-i \sigma _2-1}{\sqrt{2}}                      \\
C_{4z}^-                         & 4  & \{0,0,1\}  & \frac{7 \pi }{2} & \left\{0,0,\frac{7 \pi }{2}\right\}    & 1   & 1   & -1  & -1  & i \sigma _2  & \frac{i                                    
                                                                                                                                                       \sigma _2+1}{\sqrt{2}}                      & \frac{i \left(\sigma _2+i\right)}{\sqrt{2}}          \\
\hline                           
\widetilde{C}_{4z}^+             & 5  & \{0,0,1\}  & \frac{5 \pi }{2} & \left\{0,0,\frac{5 \pi }{2}\right\}    & 1   & 1   & -1  & -1  & -i \sigma _2 &                                            
                                                                                                                                                       \frac{i \left(\sigma _2+i\right)}{\sqrt{2}} & \frac{i \sigma _2+1}{\sqrt{2}}                       \\
\widetilde{C}_{4z}^-             & 6  & \{0,0,1\}  & \frac{3 \pi }{2} & \left\{0,0,\frac{3 \pi }{2}\right\}    & 1   & 1   & -1  & -1  & i \sigma _2  &                                            
                                                                                                                                                       \frac{-i \sigma _2-1}{\sqrt{2}}             & \frac{1-i \sigma _2}{\sqrt{2}}                       \\
\hline                           
C_{2z}                           & 7  & \{0,0,1\}  & \pi              & \{0,0,\pi \}                           & 1   & 1   & 1   & 1   & -1           & -i \sigma _2                               & i \sigma _2                                          \\
\widetilde{C}_{2z}               & 8  & \{0,0,1\}  & 3 \pi            & \{0,0,3 \pi \}                         & 1   & 1   & 1   & 1   & -1           & i \sigma _2                                & -i \sigma _2                                         \\
\hline                           
C_{2x}                           & 9  & \{1,0,0\}  & \pi              & \{0,\pi ,\pi \}                        & 1   & -1  & 1   & -1  & \sigma _3    & i \sigma _3                                & i \sigma _3                                          \\
C_{2y}                           & 10 & \{0,1,0\}  & \pi              & \{0,\pi ,0\}                           & 1   & -1  & 1   & -1  & -\sigma _3   & i \sigma _1                                & -i \sigma _1                                         \\
\widetilde{C}_{2x}               & 11 & \{1,0,0\}  & 3 \pi            & \{0,\pi ,3 \pi \}                      & 1   & -1  & 1   & -1  & \sigma _3    & -i \sigma _3                               & -i \sigma _3                                         \\
\widetilde{C}_{2y}               & 12 & \{0,1,0\}  & 3 \pi            & \{0,\pi ,2 \pi \}                      & 1   & -1  & 1   & -1  & -\sigma _3   & -i \sigma _1                               & i \sigma _1                                          \\
\hline                           
C_{2a}                           & 13 & \{1,1,0\}  & \pi              & \left\{0,\pi ,\frac{\pi }{2}\right\}   & 1   & -1  & -1  & 1   & \sigma _1    & \frac{i \left(\sigma                       
                                                                                                                                                       _1+\sigma _3\right)}{\sqrt{2}}              & \frac{i \left(\sigma _1-\sigma _3\right)}{\sqrt{2}}  \\
C_{2b}                           & 14 & \{-1,1,0\} & \pi              & \left\{0,\pi ,\frac{7 \pi }{2}\right\} & 1   & -1  & -1  & 1   & -\sigma _1   & \frac{i                                    
                                                                                                                                                       \left(\sigma _1-\sigma _3\right)}{\sqrt{2}} & \frac{i \left(\sigma _1+\sigma _3\right)}{\sqrt{2}}  \\
\widetilde{C}_{2a}               & 15 & \{1,1,0\}  & 3 \pi            & \left\{0,\pi ,\frac{5 \pi }{2}\right\} & 1   & -1  & -1  & 1   & \sigma _1    & -\frac{i                                   
                                                                                                                                                       \left(\sigma _1+\sigma _3\right)}{\sqrt{2}} & -\frac{i \left(\sigma _1-\sigma _3\right)}{\sqrt{2}} \\
\widetilde{C}_{2b}               & 16 & \{-1,1,0\} & 3 \pi            & \left\{0,\pi ,\frac{3 \pi }{2}\right\} & 1   & -1  & -1  & 1   & -\sigma _1   & -\frac{i                                   
                                                                                                                                                       \left(\sigma _1-\sigma _3\right)}{\sqrt{2}} & -\frac{i \left(\sigma _1+\sigma _3\right)}{\sqrt{2}} \\
\bottomrule                      
\end{array}                      
$                                
\end{tabular}                    
\caption{Unified elements and irreps of the single $D_4$ group and double group $^2D_{4}$. 
The rotations are represented by the axis direction $\bm n$ (which should be normalized when in use) 
and rotation angle $\omega$ about the axis (which is defined over $4\pi$).
The horizontal lines separate the elements into 7 conjugacy classes. 
The two-dimensional representation matrices are expressed in terms of Pauli matrices.  
}
\label{tab:irrep2D4}            
\end{table*}                     
\begin{table}
\begin{tabular}{c| r r r  r c c c}\toprule
$^2D_4$  & $I$  & $\widetilde{I}$ &  $2C_4$   &  $2\widetilde{C}_4$  &  $C_2+\widetilde{C}_2$   &  $2C'_2+2\widetilde{C}'_2$ & $2C''_2+2\widetilde{C}''_2$   \\
\hline
$A_{1}$  &  1  & 1       & 1                 & 1                 & 1       & 1              &   1                 \\
$A_{2}$  &  1  & 1       & 1                 & 1                 & 1       & $-1$         &  $-1$           \\
$B_1$    &  1  & 1       & $-1$            & $-1$            & 1       &  1             & $-1$            \\
$B_2$    &  1  & 1       & $-1$            & $-1$            & 1       & $-1$         & 1                \\
$ E  $     &  2  & 2       & 0                 & 0                 & $-2$  & 0              &  0             \\
\hline
$G_1$   &  2  & $-2$   & $\sqrt{2}$   &  $-\sqrt{2}$  &   0    & 0               & 0        \\
$G_2$   &  2  & $-2$   & $-\sqrt{2}$  &  $\sqrt{2}$   &   0    & 0               & 0         \\
\hline
$\omega$ & $4\pi$  &  $2\pi$  & $\pi/2$  & $5\pi/2$  & $\pi$ &  $\pi$  &   $\pi$  \\
\bottomrule
\end{tabular}
\caption{Character table for the double group $^2D_4$. Last row is the angle of rotation about the axis in each class.}
\label{tab:char2D4}
\end{table}

Table~\ref{tab:irrep2D4} summarizes all the ingredients for the elongated box.
Some discussion is in order.

\subsection{Single and double groups}

In the continuum, the full rotation group is the $SO(3)$ group; its double-cover group is the $SU(2)$ group 
which is required for the inclusion of half-integer angular momentum.
The concept of a double group can be understood by considering the 
character of the full rotation group given in Eq.~\ref{eq:charSU2}. 
One can add an extra rotation of $2\pi$ to the character  in that equation to yield the relation, 
\beq
\chi(\omega+2\pi,J)= (-1)^{2J}\chi(\omega,J). 
\eeq
We see that the extra rotation leaves the character invariant for integer $J$ as expected, 
but leads to a minus sign for half-integer $J$. 
So rotations by $4\pi$ are needed to leave the characters invariant (identity) 
for both integral and half-integral angular momentum.
This property suggests a way to construct the double group from the single group:  
by adding a new group element $\mathcal{R}$ whose role is to perform an extra $2\pi$ rotation 
to all the elements in the single group. 
It will double the number of elements (hence the name double group, or double-covered group).
The new element produced by the extra rotation on an single-group element $C_k$ will be denoted by a tilde, 
$\mathcal{R}C_k=\widetilde{C_k}$.
The number of conjugacy classes, on the other hand,  will not simply double.
The $\mathcal{R}C_3$ and $\mathcal{R}C_4$ will spawn new classes of elements not equivalent to
$C_3$ or $C_4$
because adding $2\pi$ to $2\pi/3$ and $\pi/2$ leads to new rotations, 
but $\mathcal{R}C_2$ will belong in the same class as $C_2$ because adding $2\pi$ to $\pi$ leads to the same rotations.
When applied to the $D_4$ group elements, only two new classes emerge: $\widetilde{I}$ and $\widetilde{C}_4$.
This means two new irreps in addition to the five existing ones. Their dimensions are constrained by  
$1^2 + 1^2 + 1^2 + 1^2 +2^2 + l_6^2 + l_7^2 = 16$, or $ l_6^2 + l_7^2= 8$.
The only solution is $l_6=2$, $l_7=2$. The two new two-dimensional irreps will be called $G_1$ and $G_2$.
These two even-dimensional irreps are responsible for all half-integer angular momentum in the elongated box.
Any irrep of the single group is also an irrep of the double group, with the same set of characters.
For the new irreps, the characters for the class $\mathcal{R}C_k$ are the negative of the characters of class $C_k$,
except for the $\mathcal{R}C_2$-type class for which the character is zero (when a real number is the negative of itself).
These properties lead to most of the characters in the double group. 
Additionally, the character of the $G_1$ irrep in the ${C}_3$ class is simply that of a spinor $\chi(2\pi/3,1/2)=1$, 
and in the ${C}_4$ class $\chi(\pi/2,1/2)=\sqrt{2}$.
The rest of the entries can be readily worked out by the orthogonality conditions governing  characters. 
The complete character table is given in Table~\ref{tab:char2D4}. 
It agrees with published tables (see for example Ref.~\cite{Altmann:1994}) 
(Note that the order of rows and columns in a character table does not matter).
We also checked that the multiplication table (which is a necessary closure check of the group) 
from $G_1$ agrees with that in Ref.~\cite{Altmann:1994}, 
and $G_1$ and $G_2$ have identical $16\times 16$  multiplication tables.
For most point groups, the character table can be constructed in the same manner, without knowing any representations of the group. Of course, one can readily extract the character table from the information 
given in Table~\ref{tab:irrep2D4}: for one-dimensional irreps the character is simply $1$ or $-1$; 
for two-dimensional irreps the character is the trace of the representation matrix.  
The fact that the two methods agree provides a consistency check.

It should be emphasized that Table~\ref{tab:irrep2D4} is a unified presentation for both the single group $D_4$ and double group $^2D_4$. The above discussion makes clear the relationship between the two, and how to 
construct the double group from the single one. To obtain the table just for the single group, 
simply delete the tilded rows and the last two columns ($G_1$ and $G_2$).

The full symmetry group in the elongated box must include space inversion (parity).
They are obtained by a direct product with the inversion group denoted by $C_i=\{I,i\}$ which has two elements, 
the identity and the inversion.
Therefore, for the single group $D_{4h} = D_4 \otimes C_i$, 
 and for the double group $^2D_{4h} = {^2D_4}\otimes C_i$.
 In the case of $D_{4h}$, the group elements will double to 16, with 5 new inversion-related classes.
The irreps will also double into two versions: 5 with even parity labeled by a plus sign (or g for {\it gerade}), 
5 with odd parity labeled by a minus sign (or u for {\it ungerade}). 
The character table for $D_{4h}$ can be constructed from that of $D_4$ by forming a super table $10\times 10$ consisiting 
of $2\times 2$  blocks of $5\times 5$:
adding a replica to the right and below, and its negative to the diagonal (see Ref.~\cite{Altmann:1994} for example).   
In similar fashion, the $^2D_{4h}$ group will double to 32 elements, 14 classes, and 7 even-parity and 7 odd-parity irreps. Its character table can be replicated from that of $^2D_4$ by forming a super table $14\times 14$ consisting of $2\times 2$ blocks of $7\times 7$.
In practice, however, we rarely have to work with the full content of $D_{4h}$ or $^2D_{4h}$ groups.
We can just work with  $D_{4}$  or $^2D_{4}$,  then include the consequence of space inversion 
fairly straightforwardly, {\it a posteriori}, as discussed in several places in the main text.

A useful physics consequence of this discussion is that angular momentum of both integer and half-integer values 
in the elongated box can be completely characterized by the 14 irreps of the $^2D_{4h}$ group. 
Section~\ref{sec:ang_elongated} in the main text demonstrates how it is done.

\subsection{Basis vectors in the elongated box}

The irreps of the continuum rotation group with $J = 0, 1/2, 1,3/2, \cdots$,   are defined in the
$(2J+1)$-dimensional space spanned on the basis vectors $|JM\rangle$, which are the standard spherical harmonics for integral $J$ and 
the spin spherical harmonics for half-integral $J$. These representations are reducible
under the $^2D_{4}$ group into its 7  irreps denoted by $\Gamma$.  
In other words, certain subspaces in the space spanned by $|JM\rangle$, 
are invariant under the symmetry transformations of the elongated box, furnishing irreps for
the symmetry group. To find out basis vectors corresponding to a row $\alpha$ of 
irrep $\Gamma$ we use the following projector:
\beq
P^\Gamma_\alpha = \sum_k (R^\Gamma_k)^*_{\alpha\alpha} O_k
\eeq
where $k$ runs over the group elements, $R^\Gamma_k$ is the matrix associated with rotation
$k$ in the $\Gamma$ irrep and $O_k$ is the operator that implements the rotation. For
a $|JM\rangle$ states this operator is
\beq
O_k |JM\rangle = \sum_{M'=-J}^J D^J_{MM'}(\alpha_k, \beta_k, \gamma_k) |JM'\rangle
\eeq
where $D^J_{MM'}$ is the Wigner D-matrix as a function of Euler angles $\alpha, \beta,\gamma$. 
In the case of one-dimensional irreps, $R^\Gamma_k$ are simply the characters so the 
index $\alpha$ can be dropped. 
There is freedom to choose the overall phase factor and normalization factor.
All the basis vectors are made orthonormal after they are found.

It should be pointed out that different matrix representations for the same irrep 
lead to different basis vectors.
Since equivalent matrix representations are related by a similarity transformation,
the set of basis vectors is related by the same similarity matrix.
Physics results should be independent of this ambiguity.
In the case of phaseshifts, the quantization condition involves determinants 
which are invariant under this transformation.

There is another feature in Table~\ref{tab:irrep2D4} that is worth pointing out. 
The rotations in the group elements are usually expressed as a rotation angle $\omega$ 
about a certain axis $\bm n$.  In the case of spin 1/2,
\beq
D^{1/2}(\bm n,\omega)  =
\left(
\begin{array}{cc}
 \cos\frac{\omega }{2}-i n_z \sin \frac{\omega }{2}  &-(n_y+i  n_x)  \sin \frac{\omega }{2} \\
 (n_y-i n_x) \sin \frac{\omega }{2}  & \cos \frac{\omega }{2}+i  n_z \sin \frac{\omega }{2} \\
\end{array}
\right).
\label{eq:euler1}
\eeq
%
But Euler angles are needed in the Wigner D-functions to construct the basis vectors, 
\beq
D^{1/2}(\alpha,\beta,\gamma)  =
\left(
\begin{array}{cc}
 e^{-\frac{1}{2} i (\alpha +\gamma )} \cos \frac{\beta }{2} & -e^{-\frac{1}{2} i (\alpha -\gamma )} \sin
   \frac{\beta }{2}\\
 e^{\frac{1}{2} i (\alpha -\gamma )} \sin\frac{\beta }{2} & e^{\frac{1}{2} i (\alpha +\gamma )} \cos
\frac{\beta }{2} \\
\end{array}
\right).
\label{eq:euler2}
\eeq
The traditionally-defined Euler angles are not unique 
(we use the standard active $zyz$ notation in Ref.~\cite{Tinkham:1992}). 
For example, when $\beta=0$, only the combination $\alpha+\gamma$ is  uniquely determined.
Similarly, when $\beta=\pi$, only the combination $\alpha-\gamma$ is unique.
Furthermore,  double groups require rotations of $\omega=4\pi$ to return identity for half-integral angular momentum.
The  Euler angles can be made unique by enlarging the domain of $\gamma$ from $2\pi$ to $4\pi$:
$0\le \alpha < 2\pi$,  $0\le \beta \le \pi$,  $0\le\gamma< 4\pi$, supplemented by the condition that 
$\alpha=0$ when $\beta=0$ or $\pi$ (see~\cite{Chen:2002}). 
In this way, there is an one-to-one correspondence between the two 
representations $D^{1/2}(\alpha,\beta,\gamma)$ and $D^{1/2}(\bm n,\omega)$. 
The Euler angles thus determined are given in Table~\ref{tab:irrep2D4}.
Another advantage of using the domain-extended Euler angles is in working with double groups. 
Traditionally, there is a sign ambiguity that has to be dealt with carefully, 
either by trial and error, or a factor system specially constructed to guarantee the 
one-to-one correspondence~\cite{Altmann:1994}.
Because of the unique one-to-one correspondence using the domain-extended Euler angles, 
the signs for the tilded elements are automatically and correctly produced.
For this reason, one can simply use a single-group table to represent double groups, 
thus saving a lot of space in presentation~\cite{Chen:2002}.
We choose to present the unified table of both single and double groups to make the relationship 
between the two explicit.

The basis vectors of $^2D_{4}$ are listed in Table~\ref{tab:basis2D4}.
The entire set of basis vectors can be represented by the notation
\beq
|\Gamma \alpha J l n\rangle = \sum_M C^{\Gamma \alpha n}_{J l M}  | Jl M  \rangle,
\eeq
where $\Gamma$ stands for a given irrep of the group and $\alpha$ runs from 1 to the dimension of the irrep, 
$n$ runs from 1 to $n(\Gamma,J)$, the multiplicity of $J$ in irrep $\Gamma$.
The coefficients $C^{\Gamma \alpha n}_{J l M}$ can be read off directly from the table.
These coefficients are used in Section~\ref{sec:phase_elongated} to reduce the matrix elements for phaseshifts.

\begin{table}
\begin{tabular}{c}
$      
\renewcommand{\arraystretch}{1.8}
\begin{array}{c| c | c }
\toprule
\Gamma & J     & \text{Basis in terms of  } |JM\rangle                                                             \\
\hline
A_1 & 0 & |0,0\rangle                                                                                      \\
    & 2 & |2,0\rangle                                                                                      \\
    & 4 & |4,0\rangle ;\;{1\over \sqrt{2}} ( |4,4\rangle + |4,-4\rangle)                                   \\
\hline
A_2 & 1 & |1,0\rangle                                                                                      \\
    & 3 & |3,0\rangle                                                                                      \\
    & 4 & {1\over \sqrt{2}} ( |4,4\rangle - |4,-4\rangle)                                                  \\
\hline
B_1 & 2 & {1\over \sqrt{2}} ( |2,2\rangle + |2,-2\rangle)                                                  \\
    & 3 & {1\over \sqrt{2}} ( |3,2\rangle - |3,-2\rangle)                                                  \\
    & 4 & {1\over \sqrt{2}} ( |4,2\rangle + |4,-2\rangle)                                                  \\
\hline
B_2 & 2 & {1\over \sqrt{2}} ( |2,2\rangle - |2,-2\rangle)                                                  \\
    & 3 & {1\over \sqrt{2}} ( |3,2\rangle + |3,-2\rangle)                                                  \\
    & 4 & {1\over \sqrt{2}} ( |4,2\rangle - |4,-2\rangle)                                                  \\
\hline
E   & 1 & {1\over\sqrt{2}}(|1,1\rangle \mp |1,-1\rangle)                                                   \\
    & 2 & {1\over\sqrt{2}}(|2,1\rangle \pm |2,-1\rangle)                                                   \\
    & 3 & {1\over\sqrt{2}}(|3,1\rangle \mp |3,-1\rangle); {1\over\sqrt{2}}(\mp |3,3\rangle + |3,-3\rangle) \\
    & 4 & {1\over\sqrt{2}}(|4,1\rangle \pm |4,-1\rangle); {1\over\sqrt{2}}(\pm |4,3\rangle + |4,-3\rangle) \\
\hline
G_1 & {1\over 2} & {1\over\sqrt{2}} \left(\left|{1\over 2}, {1\over 2}\right\rangle \mp \left|{1\over 2},- {1\over 2}\right\rangle\right)   \\
    & {3\over 2} & {1\over\sqrt{2}} \left(\left|{3\over 2}, {1\over 2}\right\rangle \pm \left|{3\over 2},- {1\over 2}\right\rangle\right)   \\
    & {5\over 2} & {1\over\sqrt{2}} \left(\left|{5\over 2}, {1\over 2}\right\rangle \mp \left|{5\over 2},- {1\over 2}\right\rangle\right)   \\
    & {7\over 2} & {1\over\sqrt{2}} \left(\left|{7\over 2}, {1\over 2}\right\rangle \pm \left|{7\over 2},- {1\over 2}\right\rangle\right) ;\;
                  {1\over\sqrt{2}} \left( \pm \left|{7\over 2}, {7\over 2}\right\rangle + \left|{7\over 2},- {7\over 2}\right\rangle\right) \\
\hline
G_2 & {3\over 2} & {1\over\sqrt{2}} \left(\left|{3\over 2}, {3\over 2}\right\rangle \pm \left|{3\over 2},- {3\over 2}\right\rangle\right)   \\
    & {5\over 2} & {1\over\sqrt{2}} \left(\left|{5\over 2}, {3\over 2}\right\rangle \mp \left|{5\over 2},- {3\over 2}\right\rangle\right);\;
                  {1\over\sqrt{2}} \left( \mp \left|{5\over 2}, {5\over 2}\right\rangle + \left|{5\over 2},- {5\over 2}\right\rangle\right) \\
    & {7\over 2} & {1\over\sqrt{2}} \left(\left|{7\over 2}, {3\over 2}\right\rangle \pm \left|{7\over 2},- {3\over 2}\right\rangle\right);\;
                         {1\over\sqrt{2}} \left( \pm \left|{7\over 2}, {5\over 2}\right\rangle + \left|{7\over 2},- {5\over 2}\right\rangle\right) \\
\bottomrule
\end{array}
$      
\end{tabular}
\caption{Basis vectors for the double dihedral group $^2D_{4}$ for total angular momentum up to $J=4$. 
The two-dimensional irreps ($E$, $G_1$, $G_2$) have two components indicated by upper/lower signs. 
Some irreps have two vectors for certain $J$ values (multiplicities) indicated by semicolons. 
}
\label{tab:basis2D4}
\end{table}
%

\section{Symmetry group properties in the cubic box}
\label{app:cubic}

The discussion parallels the one for the elongated box in the previous appendix.
We only outline the essential ingredients needed  in the main text.

The symmetry group of the cube box consisting of only rotations is the octahedral (or cubic) point group, 
denoted by $O$.
The $O$ group can be visualized in Fig.~\ref{fig:box_cubic}. 
The 24 operations can be divided into 5 conjugacy classes: 
the identity~($I$); 
six $\pi/2$ rotations about  the 3 axes ($6C_4$); 
three $\pi$ rotations about the 3 axes ($3C_2$);  
eight $2\pi/3$ rotations about 4 body diagonals ($8C_3$); 
and six $\pi$ rotations about axes parallel to 6 face diagonals ($6C'_2$).
The $O$ group has 5 unique irreps conventionally named $A_1$,  $A_2$,  $E$, $T_1$, and $T_2$, 
having respective dimensionality of 1, 1, 2, 3, 3 (whose squares sum to 24).   
To construct its double-covered group $^2O$, we add an extra $2\pi$ rotation to each of the 24 elements, 
which double its elements to 48.
As a result,  3 new classes emerge:  $\widetilde{I}$, $\widetilde{C}_3$ and $\widetilde{C}_4$.
This means 3 new irreps in addition to the 5 existing ones. Their dimensions are constrained by  
$1^2 + 1^2 + 2^2 + 3^2 +3^2 + l_6^2 + l_7^2 + l_8^2 = 48$, or $ l_6^2 + l_7^2 + l_8^2 = 24$.
The only solution is the combination of 3 integers (2,2,4), which can be assigned as $l_6=2$, $l_7=2$, and $l_8=4$. 
The corresponding new irreps are called $G_1$, $G_2$, and $H$. 
These three  even-dimensional irreps are responsible for all half-integer angular momentum in the cubic box.
The complete character table for the  $^2O$ is given in Table~\ref{tab:char2O}. 
The full symmetry in the cubic box must also include space inversion. The corresponding group is called 
$^2O_h$ which can be constructed by a direct product of $^2O$  with the inversion group $C_i=\{I, i\}$.
The $^2O_h$ has 96 elements and 16 irreps (8 even and 8 odd). 
The decomposition of angular momentum of both integer and half-integer values into the 16 irreps of the double group $^2O_h$ of  the cubic box  is given in Table~\ref{tab:2Oh}. 

A few words on the matrix representations in Table~\ref{tab:irrep2O}.
Operationally, $A_1$ is the identity representation.  The $T_1$ rotates the geometrical vector ($x, y, z$) whose matrices $t_k$ are  generated via $e^{-i(\bm n\cdot\bm J) \omega}$ where $(J_k)_{ij}=i\epsilon_{ijk}$, by running through the 48 elements in the order given (only distinct matrices are named). 
 Similarly, the $G_1$ rotates the spinor whose matrices  $g_k$ are generated by 
$e^{-i (\bm n\cdot\bm \sigma) \omega/2}$.
The matrices for the four-dimensional $H$ irrep $h_k$ are generated by $e^{-i (\bm n\cdot\bm J) \omega}$
where $\bm J$ are the generators of spin-3/2,
\beq
J_x=\left(
\begin{array}{cccc}
0                 & {\sqrt{3}\over 2} & 0                & 0                 \\
{\sqrt{3}\over 2} & 0                 & 1                & 0                 \\
0                 & 1                 & 0                & {\sqrt{3}\over 2} \\
0                 & 0                 &{\sqrt{3}\over 2} & 0 
\end{array}     
\right)           
,\;
J_y= \left(
\begin{array}{cccc}
0                 & -{i\sqrt{3}\over 2} & 0                & 0                 \\
{i\sqrt{3}\over 2} & 0                 & -i                & 0                 \\
0                 & i                 & 0                & -{i\sqrt{3}\over 2} \\
0                 & 0                 &{i\sqrt{3}\over 2} & 0
\end{array}   
\right)              
,\;
J_z= \left(
\begin{array}{cccc}
{3\over 2} & 0          & 0          & 0 \\
0          & {1\over 2} & 0          & 0 \\
0          & 0          & -{1\over 2} & 0 \\
0          & 0          &0           & -{3\over 2}
\end{array}
\right).
\eeq
For the remaining irreps, a sign change in the character of class $6C_4$ and $6C'_2$  connects $A_2$ to $A_1$,  $T_2$ to $T_1$, 
and $G_2$ to $G_1$, respectively.
The $E$ is a real-valued, two-dimensional irrep whose matrices can be obtained from the fact that 
it has Cartesian basis vectors $\sqrt{3}(x^2-y^2)$ and $2z^2-x^2-y^2$~\cite{Altmann:1994}. 
In other words, the $T_1$ rotations, 
which transform ($x, y, z$) to ($x', y', z'$),  will transform the basis vectors according to 
\beq
 \left(
\begin{array}{c}
\sqrt{3}(x'^2-y'^2) \\
2z'^2-x'^2-y'^2
\end{array}   
\right)   
=
\left(
\begin{array}{cc}
a_{11}  &  a_{12} \\
a_{21}  &  a_{22} 
\end{array}   
\right)    
 \left(
\begin{array}{c}
\sqrt{3}(x^2-y^2) \\
2z^2-x^2-y^2
\end{array}   
\right),      
\eeq
where the coefficients form the matrix representation for the $E$ irrep.  
The five distinct matrices thus obtained in Table~\ref{tab:irrep2O} are 
\beq
\begin{array}{ccccc}
 e_1=\left(
\begin{array}{cc}
 -1 & 0 \\
 0 & 1 \\
\end{array}
\right) & e_2=\left(
\begin{array}{cc}
 -\frac{1}{2} & -\frac{\sqrt{3}}{2} \\
 \frac{\sqrt{3}}{2} & -\frac{1}{2} \\
\end{array}
\right) & e_3=\left(
\begin{array}{cc}
 -\frac{1}{2} & \frac{\sqrt{3}}{2} \\
 -\frac{\sqrt{3}}{2} & -\frac{1}{2} \\
\end{array}
\right) & e_4=\left(
\begin{array}{cc}
 \frac{1}{2} & -\frac{\sqrt{3}}{2} \\
 -\frac{\sqrt{3}}{2} & -\frac{1}{2} \\
\end{array}
\right) & e_5=\left(
\begin{array}{cc}
 \frac{1}{2} & \frac{\sqrt{3}}{2} \\
 \frac{\sqrt{3}}{2} & -\frac{1}{2} \\
\end{array}
\right). 
\end{array}
\eeq
We also checked that $G_1$, $G_2$, and $H$ have identical $48\times 48$  multiplication tables.

It is worth emphasizing that the representation matrices for the multi-dimensional irreps 
generated by running through the 48 elements in the given order automatically acquire the correct signs for 
both single and double groups. The rotation axis $\bm n$ and the angle $\omega$ have one-to-one correspondence to the domain-extended Euler angles, as discussed in the previous appendix. The basis vectors from the Euler angles 
in the cubic box are given in Table~\ref{tab:basis2O}.

\begin{table*}                   
\begin{tabular}{c}               
$                                
\renewcommand{\arraystretch}{1.2}
\arraycolsep=6pt
\begin{array}{c | c | c | c | c | r | r | r | r | r | r | r | r}
\toprule
k  & \text{Elem}            & \bm n       & \omega            & \{\alpha,\beta,\gamma\}                                         & A_1 & A_2 & E   & T_1    & T_2     & G_1     & G_2     & H       \\
\hline
1  & \text{I}               & \{0,0,1\}   & 4 \pi             & \{0,0,0\}                                                       & 1   & 1   & \iden   & t_1    & t_1     & g_1     & g_1     & h_1     \\
\hline   
2  & \widetilde{I}              & \{0,0,1\}   & 2 \pi             & \{0,0,2 \pi \}                                                  & 1   & 1   & \iden   & t_1    & t_1     & -g_1    & -g_1    & -h_1    \\
\hline   
3  & C_{2x}                 & \{1,0,0\}   & \pi               & \{0,\pi ,\pi \}                                                 & 1   & 1   & \iden   & t_2    & t_2     & g_2     & g_2     & h_2     \\
4  & C_{2y}                 & \{0,1,0\}   & \pi               & \{0,\pi ,0\}                                                    & 1   & 1   & \iden   & t_3    & t_3     & g_3     & g_3     & h_3     \\
5  & C_{2z}                 & \{0,0,1\}   & \pi               & \{0,0,\pi \}                                                    & 1   & 1   & \iden   & t_4    & t_4     & g_4     & g_4     & h_4     \\
6  & \widetilde{C}_{2x}         & \{1,0,0\}   & 3 \pi             & \{0,\pi ,3 \pi \}                                               & 1   & 1   & \iden   & t_2    & t_2     & -g_2    & -g_2    & -h_2    \\
7  & \widetilde{C}_{2y}         & \{0,1,0\}   & 3 \pi             & \{0,\pi ,2 \pi \}                                               & 1   & 1   & \iden   & t_3    & t_3     & -g_3    & -g_3    & -h_3    \\
8  & \widetilde{C}_{2z}         & \{0,0,1\}   & 3 \pi             & \{0,0,3 \pi \}                                                  & 1   & 1   & \iden   & t_4    & t_4     & -g_4    & -g_4    & -h_4    \\
\hline   
9  & C_{31}^+               & \{1,1,1\}   & \frac{2 \pi }{3}  & \left\{0,\frac{\pi }{2},\frac{\pi }{2}\right\}                  & 1   & 1   & e_3 & t_5    & t_5     &         
                                                                                                                                                                      g_5      & g_5     & h_5     \\
10 & C_{32}^+               & \{-1,-1,1\} & \frac{2 \pi }{3}  & \left\{\pi ,\frac{\pi }{2},\frac{7 \pi }{2}\right\}             & 1   & 1   & e_3 & t_6    &         
                                                                                                                                                            t_6      & g_6     & g_6     & h_6     \\
11 & C_{33}^+               & \{1,-1,-1\} & \frac{2 \pi }{3}  & \left\{\pi ,\frac{\pi }{2},\frac{5 \pi }{2}\right\}             & 1   & 1   & e_3 & t_7    &         
                                                                                                                                                            t_7      & g_7     & g_7     & h_7     \\
12 & C_{34}^+               & \{-1,1,-1\} & \frac{2 \pi }{3}  & \left\{0,\frac{\pi }{2},\frac{7 \pi }{2}\right\}                & 1   & 1   & e_3 & t_8    & t_8     
                                                                                                                                                                     & g_8     & g_8     & h_8     \\
13 & C_{31}^-               & \{1,1,1\}   & \frac{10 \pi }{3} & \left\{\frac{\pi }{2},\frac{\pi }{2},3 \pi \right\}             & 1   & 1   & e_2 & t_9    &         
                                                                                                                                                            t_9      & g_9     & g_9     & h_9     \\
14 & C_{32}^-               & \{-1,-1,1\} & \frac{10 \pi }{3} & \left\{\frac{3 \pi }{2},\frac{\pi }{2},2 \pi \right\}           & 1   & 1   & e_2 &        
                                                                                                                                                   t_{10}  & t_{10}  & g_{10}  & g_{10}  & h_{10}  \\
15 & C_{33}^-               & \{1,-1,-1\} & \frac{10 \pi }{3} & \left\{\frac{\pi }{2},\frac{\pi }{2},0\right\}                  & 1   & 1   & e_2 & t_{11} &         
                                                                                                                                                            t_{11}   & g_{11}  & g_{11}  & h_{11}  \\
16 & C_{34}^-               & \{-1,1,-1\} & \frac{10 \pi }{3} & \left\{\frac{3 \pi }{2},\frac{\pi }{2},3 \pi \right\}           & 1   & 1   & e_2 &        
                                                                                                                                                   t_{12}  & t_{12}  & g_{12}  & g_{12}  & h_{12}  \\
\hline                                                                                                                                               
17 & \widetilde{C}_{31}^+       & \{1,1,1\}   & \frac{8 \pi }{3}  & \left\{0,\frac{\pi }{2},\frac{5 \pi }{2}\right\}                & 1   & 1   & e_3 & t_5    
                                                                                                                                                           & t_5     & -g_5    & -g_5    & -h_5    \\
18 & \widetilde{C}_{32}^+       & \{-1,-1,1\} & \frac{8 \pi }{3}  & \left\{\pi ,\frac{\pi }{2},\frac{3 \pi }{2}\right\}             & 1   & 1   & e_3 
                                                                                                                                                  & t_6    & t_6     & -g_6    & -g_6    & -h_6    \\
19 & \widetilde{C}_{33}^+       & \{1,-1,-1\} & \frac{8 \pi }{3}  & \left\{\pi ,\frac{\pi }{2},\frac{\pi }{2}\right\}               & 1   & 1   & e_3 &        
                                                                                                                                                   t_7     & t_7     & -g_7    & -g_7    & -h_7    \\
20 & \widetilde{C}_{34}^+       & \{-1,1,-1\} & \frac{8 \pi }{3}  & \left\{0,\frac{\pi }{2},\frac{3 \pi }{2}\right\}                & 1   & 1   & e_3 &        
                                                                                                                                                   t_8     & t_8     & -g_8    & -g_8    & -h_8    \\
21 & \widetilde{C}_{31}^-       & \{1,1,1\}   & \frac{4 \pi }{3}  & \left\{\frac{\pi }{2},\frac{\pi }{2},\pi \right\}               & 1   & 1   & e_2 &        
                                                                                                                                                   t_9     & t_9     & -g_9    & -g_9    & -h_9    \\
22 & \widetilde{C}_{32}^-       & \{-1,-1,1\} & \frac{4 \pi }{3}  & \left\{\frac{3 \pi }{2},\frac{\pi }{2},0\right\}                & 1   & 1   & e_2 &        
                                                                                                                                                   t_{10}  & t_{10}  & -g_{10} & -g_{10} & -h_{10} \\
23 & \widetilde{C}_{33}^-       & \{1,-1,-1\} & \frac{4 \pi }{3}  & \left\{\frac{\pi }{2},\frac{\pi }{2},2 \pi \right\}             & 1   & 1   & e_2 
                                                                                                                                                  & t_{11} & t_{11}  & -g_{11} & -g_{11} & -h_{11} \\
24 & \widetilde{C}_{34}^-       & \{-1,1,-1\} & \frac{4 \pi }{3}  & \left\{\frac{3 \pi }{2},\frac{\pi }{2},\pi \right\}             & 1   & 1   & e_2 
                                                                                                                                                  & t_{12} & t_{12}  & -g_{12} & -g_{12} & -h_{12} \\
\hline                                                                                                                                                  
25 & C_{4x}^+               & \{1,0,0\}   & \frac{\pi }{2}    & \left\{\frac{3 \pi }{2},\frac{\pi }{2},\frac{5 \pi }{2}\right\} & 1   & -1  & e_4 
                                                                                                                                                  & t_{13} & -t_{13} & g_{13}  & -g_{13} & h_{13}  \\
26 & C_{4y}^+               & \{0,1,0\}   & \frac{\pi }{2}    & \left\{0,\frac{\pi }{2},0\right\}                               & 1   & -1  & e_5 & t_{14} & -t_{14} & g_{14}  &         
                                                                                                                                                                                -g_{14}  & h_{14}  \\
27 & C_{4z}^+               & \{0,0,1\}   & \frac{\pi }{2}    & \left\{0,0,\frac{\pi }{2}\right\}                               & 1   & -1  & e_1 & t_{15} & -t_{15} & g_{15}  &         
                                                                                                                                                                                -g_{15}  & h_{15}  \\
28 & C_{4x}^-               & \{1,0,0\}   & \frac{7 \pi }{2}  & \left\{\frac{\pi }{2},\frac{\pi }{2},\frac{7 \pi }{2}\right\}   & 1   & -1  & e_4 
                                                                                                                                                  & t_{16} & -t_{16} & g_{16}  & -g_{16} & h_{16}  \\
29 & C_{4y}^-               & \{0,1,0\}   & \frac{7 \pi }{2}  & \left\{\pi ,\frac{\pi }{2},3 \pi \right\}                       & 1   & -1  & e_5 & t_{17} & -t_{17} 
                                                                                                                                                                     & g_{17}  & -g_{17} & h_{17}  \\
30 & C_{4z}^-               & \{0,0,1\}   & \frac{7 \pi }{2}  & \left\{0,0,\frac{7 \pi }{2}\right\}                             & 1   & -1  & e_1 & t_{18} & -t_{18} &         
                                                                                                                                                                      g_{18}   & -g_{18} & h_{18}  \\
\hline                                                                                                                                                                      
31 & \widetilde{C}_{4x}^+       & \{1,0,0\}   & \frac{5 \pi }{2}  & \left\{\frac{3 \pi }{2},\frac{\pi }{2},\frac{\pi }{2}\right\}   & 1   &     
                                                                                                                                       -1   & e_4 & t_{13} & -t_{13} & -g_{13} & g_{13}  & -h_{13} \\
32 & \widetilde{C}_{4y}^+       & \{0,1,0\}   & \frac{5 \pi }{2}  & \left\{0,\frac{\pi }{2},2 \pi \right\}                          & 1   & -1  & e_5 & t_{14} &         
                                                                                                                                                            -t_{14}  & -g_{14} & g_{14}  & -h_{14} \\
33 & \widetilde{C}_{4z}^+       & \{0,0,1\}   & \frac{5 \pi }{2}  & \left\{0,0,\frac{5 \pi }{2}\right\}                             & 1   & -1  & e_1 & t_{15} &         
                                                                                                                                                            -t_{15}  & -g_{15} & g_{15}  & -h_{15} \\
34 & \widetilde{C}_{4x}^-       & \{1,0,0\}   & \frac{3 \pi }{2}  & \left\{\frac{\pi }{2},\frac{\pi }{2},\frac{3 \pi }{2}\right\}   & 1   &     
                                                                                                                                       -1   & e_4 & t_{16} & -t_{16} & -g_{16} & g_{16}  & -h_{16} \\
35 & \widetilde{C}_{4y}^-       & \{0,1,0\}   & \frac{3 \pi }{2}  & \left\{\pi ,\frac{\pi }{2},\pi \right\}                         & 1   & -1  & e_5 & t_{17} &         
                                                                                                                                                            -t_{17}  & -g_{17} & g_{17}  & -h_{17} \\
36 & \widetilde{C}_{4z}^-       & \{0,0,1\}   & \frac{3 \pi }{2}  & \left\{0,0,\frac{3 \pi }{2}\right\}                             & 1   & -1  & e_1 & t_{18} &         
                                                                                                                                                            -t_{18}  & -g_{18} & g_{18}  & -h_{18} \\
\hline                                                                                                                                                            
37 & C_{2a}                 & \{1,1,0\}   & \pi               & \left\{0,\pi ,\frac{\pi }{2}\right\}                            & 1   & -1  & e_1 & t_{19} & -t_{19} & g_{19}  & -g_{19} 
                                                                                                                                                                                         & h_{19}  \\
38 & C_{2b}                 & \{-1,1,0\}  & \pi               & \left\{0,\pi ,\frac{7 \pi }{2}\right\}                          & 1   & -1  & e_1 & t_{20} & -t_{20} & g_{20}  &         
                                                                                                                                                                                -g_{20}  & h_{20}  \\
39 & C_{2c}                 & \{1,0,1\}   & \pi               & \left\{0,\frac{\pi }{2},\pi \right\}                            & 1   & -1  & e_5 & t_{21} & -t_{21} & g_{21}  & -g_{21} 
                                                                                                                                                                                         & h_{21}  \\
40 & C_{2d}                 & \{0,1,1\}   & \pi               & \left\{\frac{\pi }{2},\frac{\pi }{2},\frac{\pi }{2}\right\}     & 1   & -1  & e_4 & t_{22} &         
                                                                                                                                                            -t_{22}  & g_{22}  & -g_{22} & h_{22}  \\
41 & C_{2e}                 & \{-1,0,1\}  & \pi               & \left\{\pi ,\frac{\pi }{2},0\right\}                            & 1   & -1  & e_5 & t_{23} & -t_{23} & g_{23}  & -g_{23} 
                                                                                                                                                                                         & h_{23}  \\
42 & C_{2f}                 & \{0,-1,1\}  & \pi               & \left\{\frac{3 \pi }{2},\frac{\pi }{2},\frac{7 \pi }{2}\right\} & 1   & -1  & e_4 & t_{24} &         
                                                                                                                                                            -t_{24}  & g_{24}  & -g_{24} & h_{24}  \\
43 & \widetilde{C}_{2a}         & \{1,1,0\}   & 3 \pi             & \left\{0,\pi ,\frac{5 \pi }{2}\right\}                          & 1   & -1  & e_1 & t_{19} & -t_{19} &         
                                                                                                                                                                      -g_{19}  & g_{19}  & -h_{19} \\
44 & \widetilde{C}_{2b}         & \{-1,1,0\}  & 3 \pi             & \left\{0,\pi ,\frac{3 \pi }{2}\right\}                          & 1   & -1  & e_1 & t_{20} & -t_{20} &         
                                                                                                                                                                      -g_{20}  & g_{20}  & -h_{20} \\
45 & \widetilde{C}_{2c} & \{1,0,1\}   & 3 \pi             & \left\{0,\frac{\pi }{2},3 \pi \right\}                          & 1   & -1  & e_5 & t_{21} & -t_{21} 
                                                                                                                                                                     & -g_{21} & g_{21}  & -h_{21} \\
46 & \widetilde{C}_{2d}         & \{0,1,1\}   & 3 \pi             & \left\{\frac{\pi }{2},\frac{\pi }{2},\frac{5 \pi }{2}\right\}   & 1   & -1  & e_4 &        
                                                                                                                                                   t_{22}  & -t_{22} & -g_{22} & g_{22}  & -h_{22} \\
47 & \widetilde{C}_{2e}         & \{-1,0,1\}  & 3 \pi             & \left\{\pi ,\frac{\pi }{2},2 \pi \right\}                       & 1   & -1  & e_5 & t_{23} & -t_{23} &         
                                                                                                                                                                      -g_{23}  & g_{23}  & -h_{23} \\
48 & \widetilde{C}_{2f}         & \{0,-1,1\}  & 3 \pi             & \left\{\frac{3 \pi }{2},\frac{\pi }{2},\frac{3 \pi }{2}\right\} & 1   & -1  & e_4 
                                                                                                                                                  & t_{24} & -t_{24} & -g_{24} & g_{24}  & -h_{24} \\
\bottomrule
\end{array}
$  
\end{tabular}
\caption{Unified elements and irreps of the single cubic group $O$ and  double cubic group $^2O$. The rotations are represented by the axis direction $\bm n$ (which should be normalized) and rotation angle $\omega$
about the axis (which is defined over $4\pi$). 
The horizontal lines separate the elements into 8 conjugacy classes.}
\label{tab:irrep2O}
\end{table*}
%


%
\begin{table}
\begin{tabular}{c| c c c c c c c c}\toprule
$^2O$    & $I$    & $\widetilde{I}$ & $3C_2+3\widetilde{C}_2$& $8C_3$   & $8\widetilde{C}_3$  & $6C_4$      & $6\widetilde{C}_4$ & $6C'_2+6\widetilde{C}'_2$ \\
\hline                                                                                                                                                      
$A_{1}$  & 1      & 1               & 1                      & 1        & 1                   & 1           & 1                  & 1                         \\
$A_{2}$  & 1      & 1               & 1                      & 1        & 1                   & $-1$        & $-1$               & $-1$                      \\
$E$      & 2      & 2               & 2                      & $-1$     & $-1$                & 0           & 0                  & 0                         \\
$T_1$    & 3      & 3               & $-1$                   & 0        & 0                   & 1           & 1                  & $-1$                      \\
$T_2 $   & 3      & 3               & $-1$                   & 0        & 0                   & $-1$        & $-1$               & 1                         \\
\hline                                                                                                                                                      
$G_1$    & 2      & $-2$            & 0                      & 1        & $-1$                & $\sqrt{2}$  & $-\sqrt{2}$        & 0                         \\
$G_2$    & 2      & $-2$            & 0                      & 1        & $-1$                & $-\sqrt{2}$ & $\sqrt{2}$         & 0                         \\
$H $     & 4      & $-4$            & 0                      & $-1$     &1                    & 0           & 0                  & 0                         \\
\hline                                                                                                                                                      
$\omega$ & $4\pi$ & $2\pi$          & $\pi$                  & $2\pi/3$ & $4\pi/3$            & $\pi/2$     & $3\pi/2$           & $\pi$                     \\
\bottomrule
\end{tabular}
\caption{Character table for the double octahedral group $^2O$. Last row is the angle of rotation about the axis in each class.}
\label{tab:char2O}
\end{table}
\begin{table}
\begin{tabular}{c}
$      
\renewcommand{\arraystretch}{1.8}
\begin{array}{c| c c | l }
\toprule
\Gamma & J          & l & \text{Basis in terms of  } |JM\rangle                       \\
\hline 
A_1    & 0          &   & |0,0\rangle                                                                                \\
       & 4          &   & {\sqrt{21} \over 6} |4,0\rangle +  {\sqrt{30} \over 12} ( |4,4\rangle + |4,-4\rangle)                             \\
\hline 
A_2     & 3          &   &  {1\over \sqrt{2}} ( |3,2\rangle - |3,-2\rangle)                                            \\
\hline 
E      & 2          &  &  |2,0\rangle;\; {1\over\sqrt{2}}(|2,2\rangle + |2,-2\rangle)                  \\
         & 4          &  & {1\over\sqrt{2}}(|4,2\rangle + |4,-2\rangle);   {\sqrt{15} \over 6} |4,0\rangle - {\sqrt{42} \over 12}(|4,4\rangle + |4,-4\rangle)  \\
\hline 
T_1      & 1          &  & |1,0\rangle;\; {1\over\sqrt{2}}(|1,1\rangle - |1,-1\rangle);\;  {i\over\sqrt{2}}(|1,1\rangle + |1,-1\rangle) \\ 
       & 3         &  & |3,0\rangle;\;   {\sqrt{3} \over 4} (|3,1\rangle - |3,-1\rangle) - {\sqrt{5} \over 4}(|3,3\rangle - |3,-3\rangle);\;
      - {i\sqrt{3} \over 4} (|3,1\rangle + |3,-1\rangle) - {i\sqrt{5} \over 4}(|3,3\rangle + |3,-3\rangle)   \\ 
 & 4         &  &  -{1\over \sqrt{2}} ( |4,4\rangle - |4,-4\rangle);\; 
 - {1\over 4}(|4,3\rangle + |4,-3\rangle) - {\sqrt{7} \over 4} (|4,1\rangle + |4,-1\rangle);\;
       - {i\over 4}(|4,3\rangle - |4,-3\rangle) + {i\sqrt{7} \over 4} (|4,1\rangle - |4,-1\rangle) \\ 

\hline 
T_2      & 1          &  &-{1\over\sqrt{2}}(|2,1\rangle + |2,-1\rangle);\; -{i\over\sqrt{2}}(|2,1\rangle - |2,-1\rangle);\;  {1\over\sqrt{2}}(|2,2\rangle - |2,-2\rangle) \\ 
       & 3         &  & {1\over\sqrt{2}}(|3,2\rangle + |3,-2\rangle);\; 
         {\sqrt{5} \over 4} (|3,1\rangle - |3,-1\rangle) + {\sqrt{3} \over 4}(|3,3\rangle - |3,-3\rangle);\;
     {i\sqrt{5} \over 4} (|3,1\rangle + |3,-1\rangle)  - {i\sqrt{3} \over 4}(|3,3\rangle + |3,-3\rangle)  \\ 
 & 4         &  &  -{1\over \sqrt{2}} ( |4,2\rangle - |4,-2\rangle);\; 
  - {1 \over 4} (|4,1\rangle + |4,-1\rangle) + {\sqrt{7} \over 4}(|4,3\rangle + |4,-3\rangle);\;
     - {i \over 4} (|4,1\rangle - |4,-1\rangle)  - {i\sqrt{7}\over 4}(|4,3\rangle - |4,-3\rangle)  \\ 

\hline 
G_1    & {1\over 2} & 0, 1 &\left|{1\over 2}, {1\over 2}\right\rangle;\; \left|{1\over 2},- {1\over 2}\right\rangle   \\
       & {7\over 2} & 3, 4 & - {\sqrt{15} \over 6} \left|{7\over 2},  {7\over 2}\right\rangle -  {\sqrt{21} \over 6} \left|{7\over 2}, - {1\over 2}\right\rangle;\;
        {\sqrt{15} \over 6} \left|{7\over 2}, - {7\over 2}\right\rangle +  {\sqrt{21} \over 6} \left|{7\over 2}, {1\over 2}\right\rangle \\
 \hline 
G_2  & {5\over 2} & 2, 3 & - {\sqrt{6} \over 6} \left|{5\over 2},  {5\over 2}\right\rangle + {\sqrt{30} \over 6} \left|{5\over 2}, - {3\over 2}\right\rangle;\;
- {\sqrt{6} \over 6} \left|{5\over 2}, - {5\over 2}\right\rangle + {\sqrt{30} \over 6} \left|{5\over 2},  {3\over 2}\right\rangle  \\
       & {7\over 2} & 3, 4 &  {\sqrt{3} \over 2} \left|{7\over 2},  {5\over 2}\right\rangle - {1\over 2} \left|{7\over 2}, - {3\over 2}\right\rangle;\;
      - {\sqrt{3} \over 2} \left|{7\over 2}, - {5\over 2}\right\rangle + {1\over 2} \left|{7\over 2}, {3\over 2}\right\rangle \\
\hline 
H    & {3\over 2} & 1, 2 & \left|{3\over 2}, {3\over 2}\right\rangle;\;  \left|{3\over 2},- {3\over 2}\right\rangle;\; 
                                        \left|{3\over 2}, {1\over 2}\right\rangle;\; \left|{3\over 2},- {1\over 2}\right\rangle    \\
       & {5\over 2} & 2, 3 &  \left|{5\over 2}, {1\over 2}\right\rangle;\;  - \left|{5\over 2},- {1\over 2}\right\rangle;\;  
       {\sqrt{30} \over 6} \left|{5\over 2},  {5\over 2}\right\rangle + {\sqrt{6} \over 6} \left|{5\over 2}, - {3\over 2}\right\rangle;\;
      - {\sqrt{30} \over 6} \left|{5\over 2}, - {5\over 2}\right\rangle - {\sqrt{6} \over 6} \left|{5\over 2},  {3\over 2}\right\rangle \\
       
       & {7\over 2} & 3, 4 &   
             {\sqrt{21} \over 6} \left|{7\over 2},  {7\over 2}\right\rangle -  {\sqrt{15} \over 6} \left|{7\over 2}, - {1\over 2}\right\rangle;\;
       {\sqrt{21} \over 6} \left|{7\over 2}, - {7\over 2}\right\rangle -  {\sqrt{15} \over 6} \left|{7\over 2},  {1\over 2}\right\rangle;\;  
       {1 \over 2} \left|{7\over 2},  {5\over 2}\right\rangle + {\sqrt{3}\over 2} \left|{7\over 2}, - {3\over 2}\right\rangle;\;
       {1 \over 2} \left|{7\over 2}, - {5\over 2}\right\rangle + {\sqrt{3}\over 2} \left|{7\over 2},  {3\over 2}\right\rangle;\; \\
                                                                  
\bottomrule
\end{array}
$      
\end{tabular}
\caption{Basis vectors of  the double cubic group $^2O$ for total angular momentum up to $J=4$. 
The components for multi-dimensional irreps are indicated by semicolons. There are no multiplicities up to this $J$ cutoff.}
\label{tab:basis2O}
\end{table}
\begin{table}
\begin{tabular}{c}
$      
\renewcommand{\arraystretch}{1.8}
\begin{array}{c | c | l | l }
\toprule
\Gamma & J & \text{Basis } |JM\rangle \text{ (even branch) }    & \text{Basis } |JM\rangle  \text{ (odd branch) }                         \\
\hline
A_1 & 0 & |0,0\rangle                                                     &                                                                 \\
    & 1 &                                                                 & |1,0\rangle                                                     \\
    & 2 & |2,0\rangle                                                     &                                                                 \\
    & 3 &                                                                 & |3,0\rangle                                                     \\
    & 4 & |4,0\rangle ;\; {1\over \sqrt{2}} ( |4,4\rangle + |4,-4\rangle) &                  \\
\hline
A_2    & 4 & {1\over \sqrt{2}} ( |4,4\rangle - |4,-4\rangle)                 & \\
\hline
B_1 & 2 & {1\over \sqrt{2}} ( |2,2\rangle + |2,-2\rangle)                 &                 \\
    & 3 &                & {1\over \sqrt{2}} ( |3,2\rangle + |3,-2\rangle)                 \\
    & 4 & {1\over \sqrt{2}} ( |4,2\rangle + |4,-2\rangle)                 &                  \\
\hline
B_2 & 2 & {1\over \sqrt{2}} ( |2,2\rangle - |2,-2\rangle)                 &                \\
    & 3 &             & {1\over \sqrt{2}} ( |3,2\rangle - |3,-2\rangle)                 \\
    & 4 & {1\over \sqrt{2}} ( |4,2\rangle - |4,-2\rangle)                 &                \\
\hline                                                                                                                                                                                                        
E & 1 &       & {1\over\sqrt{2}}(|1,1\rangle \pm |1,-1\rangle)                                                   \\
  & 2 & {1\over\sqrt{2}}(|2,1\rangle \pm |2,-1\rangle)                                                   &   \\
  & 3 & & {1\over\sqrt{2}}(|3,1\rangle \pm |3,-1\rangle); {1\over\sqrt{2}}(\pm |3,3\rangle + |3,-3\rangle) \\
  & 4 & {1\over\sqrt{2}}(|4,1\rangle \pm |4,-1\rangle); {1\over\sqrt{2}}(\pm |4,3\rangle + |4,-3\rangle) & \\
\hline
G_1 & {1\over 2} & {1\over\sqrt{2}} \left(\left|{1\over 2}, {1\over 2}\right\rangle \mp \left|{1\over 2},- {1\over 2}\right\rangle\right)      & {1\over\sqrt{2}} \left(\left|{1\over 2}, {1\over 2}\right\rangle \pm \left|{1\over 2},- {1\over 2}\right\rangle\right)      \\
    & {3\over 2} & {1\over\sqrt{2}} \left(\left|{3\over 2}, {1\over 2}\right\rangle \pm \left|{3\over 2},- {1\over 2}\right\rangle\right)      & {1\over\sqrt{2}} \left(\left|{3\over 2}, {1\over 2}\right\rangle \mp \left|{3\over 2},- {1\over 2}\right\rangle\right)      \\
    & {5\over 2} & {1\over\sqrt{2}} \left(\left|{5\over 2}, {1\over 2}\right\rangle \mp \left|{5\over 2},- {1\over 2}\right\rangle\right)      & {1\over\sqrt{2}} \left(\left|{5\over 2}, {1\over 2}\right\rangle \pm \left|{5\over 2},- {1\over 2}\right\rangle\right)      \\
   & {7\over 2} &  {1\over\sqrt{2}} \left(\left|{7\over 2}, {1\over 2}\right\rangle \pm \left|{7\over 2},- {1\over 2}\right\rangle\right) ;\;
                                        {1\over\sqrt{2}} \left( \pm \left|{7\over 2}, {7\over 2}\right\rangle + \left|{7\over 2},- {7\over 2}\right\rangle\right) 
                       &  {1\over\sqrt{2}} \left(\left|{7\over 2}, {1\over 2}\right\rangle \mp \left|{7\over 2},- {1\over 2}\right\rangle\right) ;\;
                                        {1\over\sqrt{2}} \left( \mp \left|{7\over 2}, {7\over 2}\right\rangle + \left|{7\over 2},- {7\over 2}\right\rangle\right)  \\
\hline                                                                                                                                                                                                                                                                       
G_2 & {3\over 2} & {1\over\sqrt{2}} \left(\left|{3\over 2}, {3\over 2}\right\rangle \pm \left|{3\over 2},- {3\over 2}\right\rangle\right)      & {1\over\sqrt{2}} \left(\left|{3\over 2}, {3\over 2}\right\rangle \mp \left|{3\over 2},- {3\over 2}\right\rangle\right)       \\
       & {5\over 2}  &  {1\over\sqrt{2}} \left(\left|{5\over 2}, {3\over 2}\right\rangle \mp \left|{5\over 2},- {3\over 2}\right\rangle\right);\;
                                        {1\over\sqrt{2}} \left( \mp \left|{5\over 2}, {5\over 2}\right\rangle + \left|{5\over 2},- {5\over 2}\right\rangle\right) 
                            &  {1\over\sqrt{2}} \left(\left|{5\over 2}, {3\over 2}\right\rangle \pm \left|{5\over 2},- {3\over 2}\right\rangle\right);\;
                                        {1\over\sqrt{2}} \left( \pm \left|{5\over 2}, {5\over 2}\right\rangle + \left|{5\over 2},- {5\over 2}\right\rangle\right) \\
       & {7\over 2}  &  {1\over\sqrt{2}} \left(\left|{7\over 2}, {3\over 2}\right\rangle \pm \left|{7\over 2},- {3\over 2}\right\rangle\right);\;
                                        {1\over\sqrt{2}} \left( \pm \left|{7\over 2}, {5\over 2}\right\rangle + \left|{7\over 2},- {5\over 2}\right\rangle\right)
                            &  {1\over\sqrt{2}} \left(\left|{7\over 2}, {3\over 2}\right\rangle \mp \left|{7\over 2},- {3\over 2}\right\rangle\right);\;
                                        {1\over\sqrt{2}} \left( \mp \left|{7\over 2}, {5\over 2}\right\rangle + \left|{7\over 2},- {5\over 2}\right\rangle\right) \\
\bottomrule
\end{array}
$      
\end{tabular}
\caption{Basis vectors for the $^2C_{4v}$ group for total angular momentum up to $J=4$. 
The two-dimensional irreps ($E$, $G_1$, $G_2$) have two components indicated by upper/lower signs.
Some irreps have two vectors for certain $J$ values (multiplicities) indicated by semicolomns. 
 }
\label{tab:basis2C4v}
\end{table}
%

\section{Basis vectors for boosting ($^2C_{4v}$ group) }
\label{app:boost}

Boosting singles out a special direction so the group symmetry is reduced. For boosts in the z-direction, the symmetry is reduced from the $O_h$ group to 
its subgroup $C_{4v}$ (the little group), or from its double cover $^2O_h$ to $^2C_{4v}$ when half-integral angular momentum is involved. 
The $C_{4v}$ group has eight elements, divided into five conjugacy classes: 
the identity (I), two $\pi/2$ rotations about the z-axis ($2 C_4$), 
one $\pi$ rotation about the $z$-axis ($C_2$),  two mirror reflections about $xz$ and $yz$ planes ($2\sigma_v$), 
and two mirror reflections about the two diagonal planes containing the z-axis  ($2\sigma_d$). 
One can identify these operations from those of the full cubic group (both single $O_h$ and double $^2O_h$) using Table~\ref{tab:irrep2O}. 
Note that this table only displays elements that are proper rotations for both $O$ and $^2O$. 
The full symmetry groups $O_h$ and $^2O_h$ include the inversion (sometimes called improper rotations).
For $C_{4v}$, the 8 operations are  $1(I), 3 (iC_{2x}=\sigma_{v1}), 4 (iC_{2y}=\sigma_{v2}), 5 (C_{2z}), 27 (C^+_{4z}), 30 (C^-_{4z}), 37 (iC^+_{2a}=\sigma_{d1}), 38 (iC^+_{2b}=\sigma_{d2})$ 
where we have used the label $i$ to indicate the inversion and give its equivalent names.
For the double group $^2C_{4v}$, the 16 equivalent operations are  $1,2,3,4,5,6,7,8,27,30,33,36,37,38,43,44$. 
If we call these operations $S_k$, they all satisfy $S_k \bm d = \bm d$; that is, they preserve the boost direction, which is a general requirement for 
scattering between unequal-mass particles. 
For equal masses, the inversion symmetry is restored,
so that the boundary conditions are the same for $\pm\bm d$. The little group is enlarged so
that the elements obey $S_k \bm d =\pm \bm d$ and the group can be factorized into a direct product
of proper transformation and the inversion group.

The $C_{4v}$ group has five irreps conventionally called $A_1, A_2, E, B_1, B_2$ with respective dimensions $1,1,2,1,1$.
Its double cover group $^2C_{4v}$ has seven irreps called $A_1, A_2, E, B_1, B_2, G_1, G_2$ with respective dimensions $1,1,2,1,1,2,2$.
It turns out that the $^2C_{4v}$ group is isomorphic to the $^2D_4$ group (same for the single version $C_{4v}$ to $D_4$); both belong in the dihedral group family~\cite{Chen:2002}.  They have the same irreps and characters, thus the same angular momentum content. 
The difference is they have different basis vectors. 
The $^2D_4$ group consists only of proper rotations. 
The $^2C_{4v}$ group has elements that involve parity transformations. 
Only proper rotations about the $z$-axis survive in $^2C_{4v}$, while proper rotations about 
all three axes ($x,y,z$) are present in $^2D_4$. The rotations about $x$ and $y$ axis change parity. 
Consequently,  the $A_1$ irrep of $^2C_{4v}$ group  (which is the `identity' irrep of the group)
couples to all possible values of $l$, but the $A_2$ irrep only couples to even values starting at $J=4$. 
The full basis vectors for the $^2C_{4v}$ group up to $J=4$ are given in Table~\ref{tab:basis2C4v}. 
We separate them into 
two branches, one for even $l$, one for odd $l$, but they should be used as one combined basis. 
The $l$ values are implicit: for integer irreps $l=J$ (assuming two spin-0 mesons); 
for half-integer irreps $l=J\pm1/2$. So the basis vectors for $G_1$ and $G_2$ in the two branches are distinct, corresponding to two $l$ values (one even, one odd).
When the full basis is used in the reduction of phaseshifts, the results in 
Table~\ref{tab:me2C4v1} and Table~\ref{tab:me2C4v2}  are obtained.

\end{widetext}

\newpage
\bibliography{refs_group}

\end{document}